\documentclass[12pt]{article}
\title{Periodic outgassing as a result of unsteady convection in Ray Lava Lake, Mount Erebus, Antarctica}
\usepackage{authblk}
\author[a,b,*]{Janine Birnbaum}
\author[b]{Tobias Keller}
\author[b]{Jenny Suckale}
\author[a]{Einat Lev}
\affil[a]{ Lamont-Doherty Earth Observatory, Columbia University, NY, USA}
\affil[b]{Stanford University, CA, USA}
\affil[*]{Corresponding Author: janineb@ldeo.columbia.edu}

\usepackage[letterpaper, portrait, margin=1in]{geometry}
\usepackage[utf8]{inputenc}
\usepackage{pdflscape}
\usepackage{multicol}
\usepackage{graphicx}
\usepackage{amsmath}
\usepackage{amssymb}
\usepackage{gensymb}
\usepackage{setspace}
\usepackage{parskip}
\usepackage{lineno}
\usepackage{color}
\usepackage[colorinlistoftodos]{todonotes}
\usepackage{natbib}

\newcommand{\vel}{\mathbf{v}}
\newcommand{\Grad}{\boldsymbol{\nabla}}
\newcommand{\Div}{\boldsymbol{\nabla} \cdot}
\newcommand{\Dev}[1]{\underline{\mathbf{D}}(#1)}
\newcommand{\gvec}{\mathbf{g}}

\newcommand{\Gphi}{\Gamma_\phi}
\newcommand{\Gx}{\Gamma_x}
\newcommand{\GT}{\Gamma_T}

\definecolor{mygray}{gray}{0.6}
\newcommand{\e}[1]{\ensuremath{\times 10^{#1}}}

\onehalfspace

\begin{document}

\maketitle


\begin{abstract}
Persistently active lava lakes show continuous outgassing and open convection over years to decades. Ray Lake, the lava lake at Mount Erebus, Ross Island, Antarctica, maintains long-term, near steady-state behavior in temperature, heat flux, gas flux, lake level, and composition. This activity is superposed by periodic small pulses of gas and hot magma every 5-18 minutes and disrupted by sporadic Strombolian eruptions. The periodic pulses have been attributed to a variety of potential processes including unstable bidirectional flow in the conduit feeding the lake. In contrast to hypotheses invoking a conduit source for the observed periodicity, we test the hypothesis that the behavior could be the result of dynamics within the lake itself, independent of periodic influx from the conduit. We perform numerical simulations of convection in Ray Lake driven by both constant and periodic inflow of gas-rich magma from the conduit to identify whether the two cases have different observational signatures at the surface. Our simulations show dripping diapirs or pulsing plumes leading to observable surface behavior with periodicities in the range of 5-20 minutes. We conclude that a convective speed faster than the inflow speed can result in periodic behavior without requiring periodicity in conduit dynamics. This finding suggests that the surface behavior of lava lakes might be less indicative of volcanic conduit processes in persistently outgassing volcanoes than previously thought, and that dynamics within the lava lake itself may modify or overprint patterns emerging from the conduit. \end{abstract}

\newpage 

\section{Introduction}
Lava lakes provide a rare opportunity for direct observation of the near-surface portion of volcanic plumbing systems. There is a potential, therefore, to use observations at lava lakes to improve our understanding of volcanic activity. Realizing this potential hinges on the ability to distinguish the effects of intra-lake processes from those deeper in the plumbing system on surface observables. The goal of this paper is to advance our ability to interpret surface data from lava lakes by isolating how different convective regimes in the lake are reflected in surface observations. \par

We focus on the case of Ray Lake on Mount Erebus, Antarctica. Similarly to other persistently active volcanoes \citep{Blackburn1976, Francis1993, Huppert2007, Palma2011, Kazahaya1994, Stevenson1998}, Erebus emits orders of magnitude more heat and gas than would be expected based on the volume of lava erupted alone. This imbalance indicates continual recirculation of lava in the conduit with volatile-rich, buoyant magma ascending and volatile-poor, denser magma descending. Compared to other lava lakes, Erebus exhibits an intermediate level of convective vigor. It lacks the organized, well-defined, rigid plates found at Kilauea and Erta Ale \citep{Harris2008, Harris2005, Patrick2016}, but neither does it show a severely disrupted surface indicating chaotic circulation like on Ambrym \citep{Carniel2003,Lev2019}. In this intermediate convective regime, we hypothesize that both conduit flow and the lava lake convection are reflected in surface observables. \par

The behavior of Ray Lake is characterized by periodic cycles of increased outgassing with a period of 5-18 min, punctuated by sporadic Strombolian eruptions \citep{Calkins2008, Oppenheimer2008, Oppenheimer2009, Peters2014, Peters2014a, Sweeney2008}. The periodic outgassing is phase locked with variations in surface velocity, temperature, heat flux, and gas composition \citep{Oppenheimer2009}. These observations suggest that the physical transport of magma controls the outgassing, but do not clarify which physical process leads to the periodicity.\par

Bidirectional flow in conduits is prone to instabilities \citep{Beckett2014, Huppert2007, Kazahaya1994, Stevenson1998, Suckale2018}, which may explain periodic signals at Mount Erebus \citep{Oppenheimer2009}. By this interpretation, the surface record is assumed to be indicative of processes occurring deeper in the conduit. It implicitly assumes that lake dynamics do not exert strong controls on surface observables. Here, we test this assumption through numerical simulations. We hypothesize that convection within the lava lake itself may lead to the observed periodicity, even in the absence of conduit-driven fluctuations. \par

We use a model of gas-buoyancy driven convection coupled to evolving vesicularity, temperature, and crystallinity-dependent rheology. First, we impose a constant inflow of bubble-rich magma from the conduit to isolate the effects of lake convection. We test a range of magma inflow, gas loss, and heat loss rates to investigate pertinent convective regimes. Secondly, we consider how lake behavior modulates periodicities imposed from the conduit. We compare model outputs of surface velocity, gas flux and heat flux to observational data. Our results reveal several regimes of lake circulation. We identify their characteristic signatures and map out their controlling parameters. We find that lake convection alone can produce periodic surface behavior similar to observations without requiring periodicity in the conduit. \par

\begin{figure}
\begin{center}
\includegraphics[width=\textwidth]{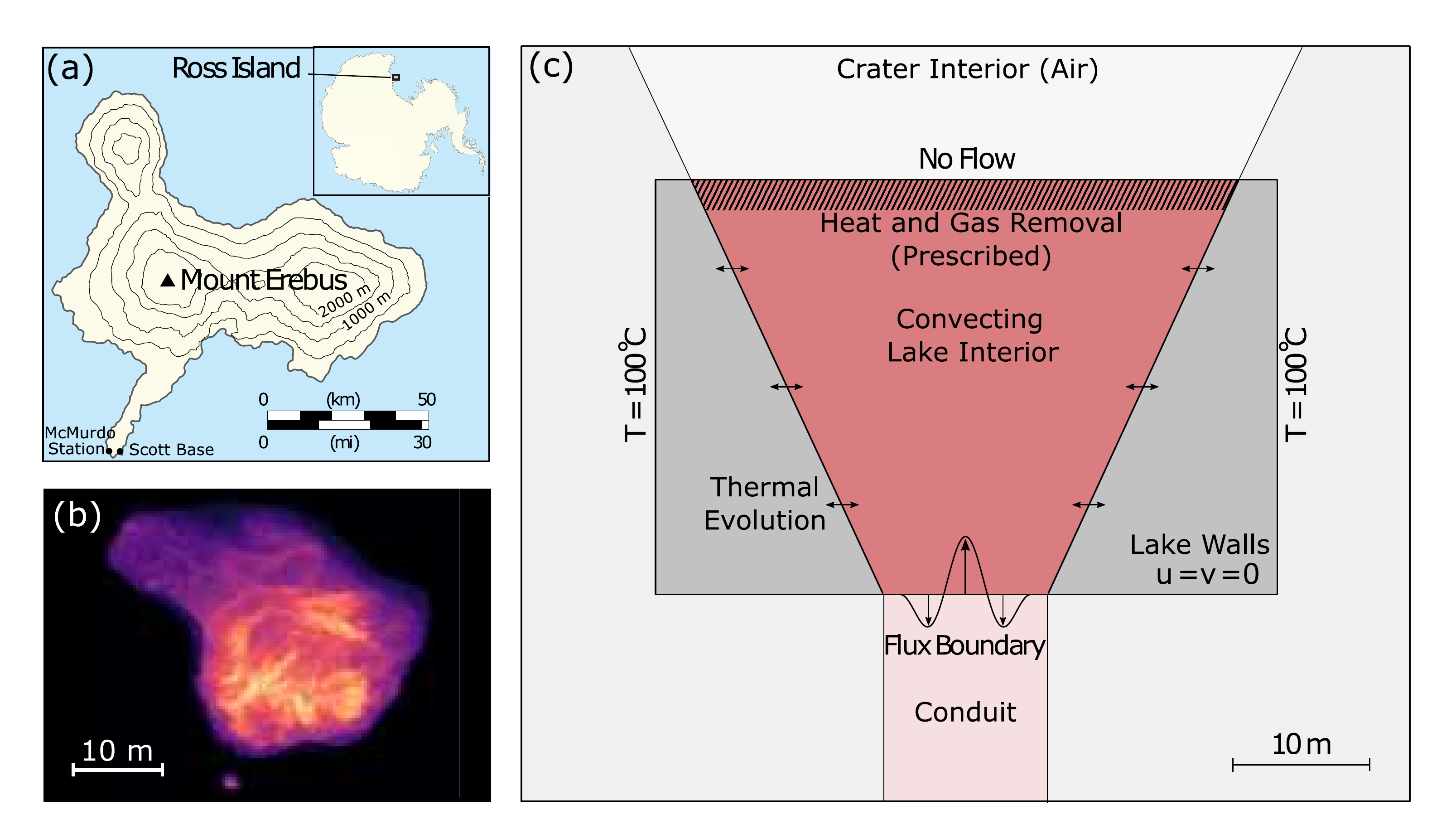}\\
\caption{(a) Map of Ross Island, Antarctica showing proximity of Mount Erebus to McMurdo Station (modified from \citet{Csatho2008}); (b) example thermal image of the lava lake surface which we use to calculate surface velocity; (c) overview of model domain highlighting boundary conditions, an imposed surface sink of gas and heat, core-annular flow from the conduit, and regions where the lake geometry evolves through thermal erosion and crystallization.}
\label{fig:Intro}
\end{center}
\end{figure}

\section{Observational constraints on Ray Lake}
Mount Erebus is part of the volcanic group on Ross Island, Antarctica. It supports a persistently active lava lake located at the bottom of the Inner Crater (Fig.~\ref{fig:Intro}). Ray Lake has been openly convecting at least since its discovery in 1972, and appears to be in near steady state \citep{Calkins2008, Dibble2008, Peters2014a}. Observational studies have employed visual, infrasound, and gas monitoring to characterize the convective and explosive activity \citep{Dibble2008, Peters2014}. Averaged over several hours, the system has approximately constant surface temperature, heat flux, and gas flux. Superimposed on this steady mean are cycles of increased outgassing, heat loss, and surface velocities with periods of 5 to 18 minutes \citep{Calkins2008, Oppenheimer2008, Oppenheimer2009, Peters2014, Peters2014a, Sweeney2008}. The steady and periodic behavior are occasionally punctuated by non-periodic Strombolian eruptions in which mild explosions disrupt the surface and partially evacuate the lake \citep{Dibble2008}. Larger explosions have revealed a simple cone-like lake geometry $\sim$40 m across and $\sim$30 m deep, with a 4-10 m wide conduit opening at the base of the lake \citep{Oppenheimer2009}.

Based on thermal imaging \citep{Calkins2008, Oppenheimer2009}, lake surface temperatures vary from $\sim 275-900 \celsius$ with mean temperatures $\sim 525-750 \celsius$ \citep{Calkins2008}. \citet{Calkins2008} found periodic increases in maximum and mean temperature of $\sim 120 \celsius$ and $\sim 20 \celsius$, respectively, every $\sim 5$ min, with thermal maxima lasting $\sim 1-2$ min. Ground-based thermal flux estimates at high lake level (surface area of $\sim$ 1400 m$^2$) are 30 $\pm$ 10 MW \citep{Calkins2008, Oppenheimer2009}. Satellite observations show higher maximum radiant fluxes up to 100 MW \citep{Wright2008}. \par

A 5-18 min periodic behavior is evident in the mean and peak surface velocities \citep{Peters2014}. Mean surface velocities are typically less than $\sim$ 0.1 m/s but can reach up to $\sim$ 0.15 m/s at high activity \citep{Oppenheimer2009, Peters2014}. Peak velocities vary between $\sim$ 0.1-0.5 m/s over periods of 5-10 minutes \citep{Oppenheimer2009} but may exceed 0.8 m/s when the lake is most active \citep{Calkins2008, Lev2019}. Strombolian eruptions are not correlated with the periodic activity and do not regulate surface velocities beyond the $\sim$5 min in which the lake refills \citep{Peters2014}. \par

Gas emission studies show a periodic total flux and composition phase-locked with surface activity \citep{Peters2014}. Plume measurements of total SO$_2$ flux show a dominant 10 min periodicity with a flux of $\sim 0.7 \pm 0.3$ kg/s SO$_2$ \citep{Sweeney2008}. Periodic peaks in surface activity are associated with emission of gas enriched in SO$_2$, H$_2$O, HCl, and HF with respect to CO$_2$ and an increase in CO$_2$/OCS \citep{Oppenheimer2008, Oppenheimer2009, Peters2014}. Calculations based on gas composition and SO$_2$ flux measurements yield an estimate for the total gas flux of $\sim$27.3 kg/s from the volcano summit, although it is uncertain what portion of the total flux derives from passive outgassing apart from lake activity. \par

\section{Methods}

We perform numerical simulations of a simplified 2D representation of Ray Lake using a mixed finite-element, finite-difference code (based on \citet{Keller2013}. The model takes a phase-averaging approach in which solid, liquid, and gas phases are described as interpenetrating continuum fields at the system scale representing volume-averaged phase interactions at the local scale \citep{Drew1983}. 

In this framework, the magma is an aggregate of liquid melt, solid crystals and vapor bubbles where the collective flow of the aggregate dominates the system-scale dynamics. The phases move collectively with the magma velocity determined by solving the Stokes equations, while bubbles are allowed to segregate vertically based on a hindered-Stokes law. The bubble fraction also diffuses as a result of local-scale fluctuations of bubble motion relative to system-scale flow \citep{Segre2001, Mucha2004}. We do not allow for segregation of crystals, which is appropriate for small grains with minimal buoyancy contrast to the carrier melt. The relatively high viscosity ($>10^4$ Pa s) does not permit significant settling over the timescale of lake convection cycles, even of the 5-10 cm anorthoclase megacrysts present at Erebus \citep{Molina2012, Moussallam2013}. Additionally, we consider crystals to be in near-equilibrium with a melt of constant composition. Thus, the main effect of crystals in the system is their stiffening effect on the aggregate viscosity, which becomes shear-thinning at high strain rates. 

\subsection{Model description}
We estimate the Reynolds number of convection in Ray Lake using
\begin{equation}
\mathrm{Re} = \frac{\rho u L}{\eta} \: ,
\end{equation}
where $\rho$ is the liquid density ($\sim$2600 kg/m$^3$), $u$ is the characteristic flow velocity ($\sim$0.2 m/s), $L$ a characteristic length we take as the conduit radius ($\sim$5 m), and $\eta$ the dynamic viscosity ($\sim$10$^4$ Pa s), yielding  $\mathrm{Re} \approx 0.26$. The low value for $\mathrm{Re}$ justifies neglecting inertial terms. \par

Mass and momentum are conserved for Stokes flow of an incompressible fluid:
\begin{subequations}
\begin{align}
    \label{eq:mass}
   \Div \vel &= 0 \: , \\
    - \Div 2 \bar{\eta} \Dev{\vel} + \Grad P &= \bar{\rho} \gvec \: ,
    \label{eq:momentum}
   \end{align}
\end{subequations}
where $\vel$ is the velocity, $\bar{\rho}$ is the density, $\bar{\eta}$ the viscosity and P the pressure in the aggregate volume, g the gravitational acceleration and $\Dev{\vel}$ the deviatoric strain rate tensor: 
\begin{equation}
    \Dev{\vel} = \dfrac{1}{2} (\Grad \vel + [\Grad \vel]^T) \: .
\end{equation}
The aggregate density $\bar{\rho}$ is the volume average of phase densities:  
\begin{equation}
 \bar{\rho} = (1-\phi)[(1-\chi) \rho_l+\chi \rho_\chi]+\phi \rho_g  , \:
\end{equation}
where $\rho_l$ is the density of the melt, $\rho_\chi$ a weighted average density of the minerals present \citep{Klein2002,Moussallam2013} and $\rho_g$ the gas density (Table \ref{tbl:RefParameters}); $\phi$ is the volume fraction of vapor in the aggregate volume, and $\chi$ the volume fraction of crystals relative to the silicate volume: 
\begin{equation}
 \chi = \frac{\phi_\chi}{\phi_\chi + \phi_l} \: ,
\end{equation}
where $\phi_\chi$ and $\phi_l$ are the volume fraction of crystals and melt per unit volume of the aggregate, respectively. \par

We couple equations \ref{eq:mass} and \ref{eq:momentum} to the conservation of energy,
\begin{equation}
	\frac{\partial T}{\partial t} = - \vel \cdot \Grad T + \kappa_T \Grad^2 T + \frac{L_\chi}{c_P} \Gamma_\chi + \GT \: ,
\end{equation}
where the specific heat capacity, $c_p$, and the thermal diffusivity, $\kappa_T$, are both assumed constant, $L_\chi$ is the latent heat of crystallization, and $\GT$ a parameterized surface cooling rate:
\begin{equation}
	\GT = -\dfrac{T-T_{atm}}{\tau_T} \exp\left(-\dfrac{z}{\delta}\right) \: ,
	\label{eq:GT}
\end{equation}
where $T_{atm}$ is the air temperature above the lake, buffered at the condensation point of water vapor (100 \celsius), $\tau_T$ is the characteristic cooling time, $z$ the depth coordinate, and $\delta$ the characteristic length of the cooling boundary layer. \par

We model the evolution of crystallinity by an advection-reaction equation
\begin{equation}
	\frac{\partial \chi}{\partial t} = -\vel \cdot \Grad \chi + \Gx \: ,
\end{equation}
where $\Gamma_\chi$ is a volumetric crystallization rate given by: 
\begin{equation}
 \Gamma_\chi = - \frac{\chi - \chi^{eq}}{\tau_\chi} \: ,
\end{equation}
with $\tau_\chi$ the characteristic time of crystallization adjusted to be rapid compared to advective transport, and $\chi^{eq}$ the equilibrium crystallinity taken as a function of temperature using a power-law fit of the form:
\begin{equation}
\chi^{eq} = \left( \dfrac{T-T_{liq}}{T_{sol}-T_{liq}} \right) ^q \: ,
\end{equation}
where $T_{sol}$ is the solidus, $T_{liq}$ the liquidus, and $q$ an exponent near unity. The fitting parameters are determined to approximate the equilibrium crystallinity with $T$ reported by \citet{Moussallam2013}.
\par

The vesicularity evolves as 
\begin{equation}
    \frac{\partial \phi}{\partial t} = -\Grad \cdot \phi \vel_g + \kappa_{\phi} \Grad^2 \phi + \Gphi \: ,
    \label{eq:gas_evolution}
\end{equation}
where $\kappa_\phi$ is the diffusivity due to local-scale fluctuations in bubble motion \citep{Mucha2004, Segre2001}, and $\vel_g$ is the gas velocity given by a hindered-Stokes segregation law: 
\begin{equation}
	\vel_g = \vel -\frac{2{a_0}^2}{9 \bar{\eta}} (1-\phi)^\mu \Delta \rho \gvec \: ,
	\label{eq:gas_vel}
\end{equation}
where $a_0$ is the average radius of bubbles or bubble clusters, $1 \leq \mu \leq 5$ is the hindering exponent \cite{Richardson1954}, which we set to 3 \citep{Manga1996}; $\Delta \rho$ is the density contrast between vapor and crystal-melt mixture. In \eqref{eq:gas_evolution}, $\Gphi$ is a parameterized outgassing rate,
\begin{equation}
	\Gphi = - \dfrac{\phi}{\tau_\phi} \exp\left(-\dfrac{z}{\delta}\right) \: , \label{eq:Gphi}
\end{equation}
in which $\tau_\phi$ is the characteristic outgassing time, and $\delta$ the same as the cooling boundary layer depth. \par

Laboratory experiments show a strong dependence of viscosity on crystal content (see \citet{Costa2009}, and refs therein). The particle-stiffening effect was described by \citet{Krieger1959} and numerous follow-up studies as recently reviewed by \citet{Mader2013} in the magmatic context. We assume a viscosity model with a smooth step increase where crystals become closely packed and form a contiguous solid (Fig.~\ref{fig:Crystallinity}): 
\begin{equation}
    \bar{\eta} =  \left[\eta_s \exp(-\lambda_s(1-\chi))\right]^{X} \times \left[\eta_\ell \exp(\lambda_l \chi)\right]^{(1-X)} , \:
    \label{eq:ClsrEta}
\end{equation}
where $\eta_l$ and $\eta_s$ are the pure melt and solid viscosities, respectively \citep{Giordano2008}, $\lambda_l$ and $\lambda_s$ are slopes of $T$-dependent viscosity away from the step, and X is a smooth step function,
\begin{equation}
    X = \dfrac{1}{2} \left[1 + \tanh\left( \dfrac{\chi-\chi_\mathrm{crit}}{w_\chi} \right)\right] , \:
    \label{eq:RheoTrans}
\end{equation}
which is centered about a critical crystallinity \citep{Arzi1978, Costa2005}, $\chi_{crit}$, and has a width of $w_\chi$ \citep{Costa2009}. We cap the maximum viscosity at 10$^{12}$ Pa s, which is below the viscosity of solid rock at these conditions but presents a large enough contrast with the lake interior to result in rigid lake walls on the time scale of interest. To limit complexity, we do not include the effects of bubbles on magma viscosity. \par

Crystal- and bubble-bearing magmas exhibit strain-rate dependent behavior \citep{Saar2001, Caricchi2007, Renner2000, Heymann2002, Costa2009, Pistone2012, Mader2013, LeLosq2015}. We impose a shear-thinning power-law rheology above $10^{-3}$ s$^{-1}$ \citep{Caricchi2007}. \par

The effects of thermal expansivity and pressure compressibility of phase materials for $\Delta T \approx 50 \celsius$, and $\Delta P \approx$ 50 kPa are small compared to the density contrasts between phases, therefore we neglect $P,T$-effects on density. We also find that the bubble segregation velocity, $\vel^v$, remains small with respect to the magma velocity field, and hence its non-zero divergence may be neglected. By neglecting both gas compressibility and the non-solenoidal flow of multi-phase segregation, we obtain an incompressible flow model. \par 

The imposed parameterized surface outgassing and cooling rates represent the collective effects of diffusive, advective, and radiative heat loss from the lake surface (eqs. \ref{eq:GT} \& \ref{eq:Gphi}). Therefore, we do not resolve small-scale processes contributing to the removal of gas and heat near the surface, which are thought to include bubble coalescence and fragmentation \citep{Blower2001}. We select the characteristic scales for outgassing and heat removal to be consistent with field observations of gas and heat flux assuming a 1400 m$^2$ surface area \citep{Calkins2008, Oppenheimer2009, Wright2008}. \par

\subsection{Dimensional analysis}
\label{sec:Dimensional_Analysis}
We identify the characteristic physical scales of the problem by performing a dimensional analysis of the governing equations. We scale variables and parameters by the following dimensional scales:
\begin{subequations}
    \begin{align}
        x = l_0 x', \ \ z = l_0 z', \ \ \vel = v_0 \vel', \ \ \vel^v = v_0 {\vel^v}', \ \ t = \frac{l_0}{v_0} t', \\
        P = \rho_0 g_0 l_0 P', \ \ T = T_0 T', \ \ \phi = \phi_0 \phi', \ \ \chi = \chi_0 \chi', \\ \eta = \eta_0 \eta', \ \ \rho = \rho_0 \rho', \ \
        \mathbf{g} = g_0 \hat{\mathbf{z}},
    \end{align}
\end{subequations}
where 
\begin{equation}
     v_0 = \frac{\rho_0 g_0 l_0^2}{\eta_0} \: ,\\
\end{equation}
is the characteristic speed of a Stokes diapir of radius $l_0$, the conduit radius, $g_0$ the acceleration due to gravity, $\rho_0$ the liquid density, $\eta_0$ the liquid viscosity at the inflow temperature, $T_0$, $\phi_0$ the inflow vesicularity, and $\chi_0$ the crystallinity at $T_0$. We substitute these scales into the governing equations and drop primes to find the dimensionless form:
\begin{subequations}
    \begin{align}
        \Grad P &= \Div \eta \Dev{\vel} + \rho \hat{\mathbf{z}} \\
        \Div \vel &= 0 \\
        \frac{\partial \phi}{\partial t} &= - \Grad \left[ \vel - \mathrm{R_{segr}} \frac{\hat{\mathbf{z}}}{\eta} \left( 1 - \phi_0 \phi \right)^\mu \right] + \frac{1}{\mathrm{Pe_\phi}} \Grad^2 \phi - \mathrm{Da_\phi} \phi \exp \left( \frac{-z}{\mathrm{d}} \right) \\
        \frac{\partial \chi}{\partial t} &= - \vel \cdot \Grad \chi + \mathrm{Da_\chi} \left( \chi - \frac{\chi^{eq}}{\chi_0} \right) \\
        \frac{\partial T}{\partial t} &= - \vel \cdot \Grad T + \frac{1}{\mathrm{Pe_T}} \Grad^2 T + \frac{\chi_0 \mathrm{Da_\chi}}{\mathrm{St}} \left( \chi - \frac{\chi^{eq}}{\chi_0} \right) - \mathrm{Da_T} \left(T - \frac{T_{atm}}{T_0} \right) \exp \left(-\frac{z}{\mathrm{d}} \right) \:.
    \end{align}
\end{subequations}
 A set of dimensionless numbers emerges from this analysis. We define $\mathrm{R_{segr}}$ as the ratio between the characteristic speeds of bubble ascent and magma convection:
 \begin{equation}
     \mathrm{R_{segr}} =  \frac{2 a_0^2 \Delta \rho g_0}{9 \eta_0 v_0} \propto \frac{a_0^2}{l_0^2} \: .
     \label{eq:R_segr}
 \end{equation}
In our models, $l_0$ is much larger than the bubble radius ($2.5$ cm  $\leq a_0 \leq 10$ cm), suggesting that magma convection should prevail over bubble segregation. \par
We find three Damkoehler numbers which compare the volumetric rates of outgassing, crystallization, and cooling to the rate of magma convection:
\begin{subequations}
    \begin{align}
         \label{eq:Da_phi}
        \mathrm{Da_\phi} &= \frac{l_0 / v_0}{\tau_\phi} \: , \\
        \mathrm{Da_\chi} &= \frac{l_0/v_0}{\tau_\chi} \: , \\
        \mathrm{Da_T} &= \frac{l_0/v_0}{\tau_T} \: .
    \end{align}
\end{subequations}
For small values of $\mathrm{Da_\phi}$, we expect outgassing to be inefficient and gas to accumulate in the domain, while large values of $\mathrm{Da_\phi}$ would result in efficient outgassing. When $\mathrm{Da_T}$ is small we expect surface cooling to dominate and a stagnant-lid regime to prevail, whereas for large $\mathrm{Da_T}$ we expect convective transport to exceed surface cooling and for the lake to remain open. Here, we use only large $\mathrm{Da_\chi}$ consistent with near-equilibrium evolution of crystallinity. \par
We find two P\'eclet numbers relating advective and diffusive transport of bubbles and heat:
\begin{subequations}
    \begin{align}
        \mathrm{Pe_\phi} &= \frac{l_0 v_0}{\kappa_\phi} \\
        \mathrm{Pe_T} &= \frac{l_0 v_0}{\kappa_T} 
    \end{align}
\end{subequations}
Both P\'eclet numbers are very large ($\mathrm{Pe_\phi} > 10^6$ and $\mathrm{Pe_T > 10^7}$) in our simulations, consistent with the advection-dominated transport of gas and heat expected for Erebus. \par
We compare the ratio of sensible heat to latent heat via the Stefan number, 
\begin{equation}
    \mathrm{St} = \frac{c_p T_0}{L_\chi} \: .
\end{equation}
We find $\mathrm{St} \approx 3.2$ for parameters of interest here and hence expect that latent heat may be important with respect to sensible heat.
\par
A final dimensionless parameter, $\mathrm{R_{in}}$, compares the imposed inflow speed at the conduit mouth to the characteristic speed of free convection:
\begin{equation}
    \mathrm{R_{in}} = \frac{u_{in}}{u_{cnvt}} = \frac{u_{in} \eta_0}{\phi_0 \Delta\rho_0 g_0 l_0^2} \: ,
\end{equation}
For high values of $\mathrm{R_{in}}$, the inflow rate exceeds the rate of convective transport, which should result in a pile-up of buoyant material near the boundary. When $\mathrm{R_{in}}$ is small, material is removed from the inlet more quickly than it is fed in, which should result in a dripping instability. \par

\subsection{Numerical model setup}

\subsubsection{Model discretization}
The velocity-pressure solution is obtained by a continuous-Galerkin (CG) finite-element method, with variables discretized on a regular mesh of rectangular elements with linear shape functions for velocity and piece-wise constant ones for pressure \citep{Brenner1994, Keller2013}. To avoid the need for stabilization in a CG-based formulation, we discretize the advection-diffusion-reaction equations for temperature, vesicularity and crystallinity by staggered-grid finite-differences on a collocated grid. \par
The model uses a second-order accurate, upwind-biased Fromm's method for grid-based advection \citep{Fromm1968}. Temperature and crystallinity are advected using the aggregate velocity only, whereas the vesicularity is advected based on the vapor velocity, which combines magma convection and bubble segregation (\eqref{eq:gas_vel}).  \par

\subsubsection{Geometry}
A novel component of our model setup is that the specific geometry of the lava lake can evolve with the flow dynamics. As illustrated in Fig.~\ref{fig:Intro}, our two-dimensional model domain is rectangular and includes the cold host rock around the lake. The surface extent of the lake and the width of the conduit mouth are initially set at 40 m and 10 m diameter, respectively, consistent with field observations of a cone-shaped lake \citep{Dibble2008, Oppenheimer2009}. We initiate simulations with an $\sim1.5$ m thick internal thermal boundary layer that linearly connects the lateral edges of the lake to the conduit. The initial temperature of the lake interior is set to the upwelling magma temperature, T = 970$\celsius$, consistent with the observed $\sim$30\% crystallinity in the lake \citep{Moussallam2013}; lake vesicularity is initially set to $\phi$ = 0 everywhere. The walls of the lake are initiated with the same, temperature-dependent material properties as the magma in the lake interior with an initial temperature of T = 100$\celsius$, assuming it is buffered by water vapor condensation. Throughout the simulation, the lake bed can dynamically evolve as a result of diffusive cooling, crystallization and thermal erosion by magma flow.\par

The side boundaries of the rectangular domain are no-slip (u = w = 0) and insulating (${\partial T}/{\partial \mathbf{n}}$ = 0 normal to boundary). Due to the relative inefficiency of thermal diffusion, the lake walls remain close to their initial T = 100$\celsius$, resulting in a high enough viscosity to keep the walls essentially rigid over a model run. The velocity field in the walls where the viscosity remains at maximum (10$^{12}$ Pa s) remains fixed at zero and is excluded from calculations for reasons of efficiency. The effect is akin to a no-slip condition along the lake bed. \par
 
Except for the conduit itself, the bottom boundary is no-slip (w = u = 0) and isothermal. Within the conduit, in- and outflow of magma is imposed by a vertical velocity of sinusoidal shape: $\sin(2 \pi x)$ for asymmetrical flow, and $(\cos(2 \pi x) + \cos(4 \pi x))/2$ for symmetrical flow. The symmetrical condition is broadly consistent with core-annular flow which has been suggested to dominate  bidirectional flow in volcanic conduits \citep{Beckett2011, Stevenson1998, Suckale2018}. The horizontal velocity remains fixed at u = 0. To avoid forced erosion of the thermal boundary layers along the conduit mouth, the velocity profile is applied on the central 8 m of the 10 m conduit only. We perform simulations with both constant and periodic influx. The latter is captured through sinusoidal variability to the inflow speed with time. \par

The top boundary is free-slip (${\partial u}/{\partial z}$ = 0, w = 0). Gas and heat are extracted from the top of the lake by the distributed sink terms $\GT$ and $\Gphi$ that decay exponentially away from the surface over the depth $\delta$ (eqs. \ref{eq:GT}, \ref{eq:Gphi}). Additionally, we allow vapor segregation across the lake surface but observe that the hindered-Stokes law alone results in a minor outgassing. Disruption of a chilled surface layer by bulging under local gas accumulation is inhibited by this technique, which biases the simulations towards a stagnant-lid regime. To keep the lake open, we select characteristic cooling times sufficiently long to prevent significant freezing of the surface during the model run time. \par

\section{Lava lake simulation results}
\subsection{Reference parameters}
Observational constraints leave considerable uncertainty in the parameters governing the dynamics of Ray Lake. We choose a set of reference parameter values and associated dimensionless numbers, summarized in Table \ref{tbl:RefParameters}, that represent our best estimate of conditions in Ray Lake. \par

\begin{table}
\begin{center}
\begin{tabular}{l c c c}
\textbf{Parameter} & \textbf{Symbol} & \textbf{Units} & \textbf{Values} \\
\hline\hline
Melt Density$^1$ & $\rho^m$ & kg/m$^3$ & 2545\\
Crystal Density$^{1,2}$ & $\rho^\chi$ & kg/m$^3$ & 2720\\
Vapor Density$^3$ & $\rho^\phi$ & kg/m$^3$ & 1\\
Thermal Conductivity$^4$ & k & W/m/K & 1.53\\
Bubble Diffusivity & $\kappa_\phi$ & m$^2$/s & 10$^{-6}$ \\
Heat of Crystallization & $L_\chi$ & J/kg & 0, \textbf{4.1}$\boldsymbol{\times 10^5}$\\
Heat Capacity$^4$ & c$_P$ & J/kg/K & 1367\\
Solidus$^1$ & T$_{sol}$ & $\celsius$ & 884.5 \\
Liquidus$^1$ & T$_{liq}$ & $\celsius$ & 1030 \\
Crystallinity Exponent & q & & 1.25 \\
Bubble Radius & $a_0$ & cm & 2.5, \textbf{5}, 10 \\
Bubble Hindering & $\mu$ & & 3 \\
Reference Melt Viscosity$^5$ & $\eta$ & Pa.s & 10$^{4}$\\
Rheology Exponent & n & & 1, \textbf{2}, 3 \\
Cohesion & c & Pa &\textbf{0},$10^7$ \\
Solid Exponential Weakening$^6$ & $\lambda_s$ & & 4.5 \\
Liquid Exponential Hardening$^5$ & $\lambda_l$ & & 3.5 \\
Critical Crystallinity$^6$ & $\chi_{crit}$ & vol & 0.6 \\
Crystallinity Step Width$^6$ & $w_\chi$ & vol & 0.2 \\
Cooling Time & $\tau_T$ & hr & 10,\textbf{15},20 \\
Outgassing Time & $\tau_\phi$ & s & 40, \textbf{80}, 160, $\infty$ \\
Outgassing Depth & $\delta$ & m & 0.5, \textbf{1}, 2 \\
Inflow Temperature & $T_0$ & $\celsius$ & 965, \textbf{970}, 975 \\
Inflow Vesicularity & $\phi_0$ & vol & 0.1, \textbf{0.2}, 0.3 \\
Inflow Velocity & $u_{in}$ & m/s & 0.1, \textbf{0.2}, 0.3, 0.4, 0.5 \\
Inflow Symmetry & & & \textbf{YES}, NO \\
Inflow Period & $\tau_{u_{in}}$ & min & 5, 10, 20, $\boldsymbol{\infty}$ \\
\hline
\textbf{Dimensionless Number} & \\
\hline \hline
Inflow/Convection & $\mathrm{R_{in}}$ && \textbf{5.0} (2.5--12)\e{-2} \\
Segregation/Convection & $\mathrm{R_{segr}}$ && \textbf{2.2} (0.56--8.9)\e{-5} \\
Gas Removal/Advection & $\mathrm{Da_\phi}$ && \textbf{3.9} (3.4--4.4)\e{-3} \\
Heat Removal/Advection & $\mathrm{Da_{T}}$ && \textbf{5.7} (4.3--8.6)\e{-6} \\
Crystal Reaction/Advection & $\mathrm{Da_{\chi}}$ && \textbf{3.1} (2.1-5.6)\e5 \\
Gas Advection/Diffusion & $\mathrm{Pe_{\phi}}$ && \textbf{8.1} (7.1--9.1)\e6 \\
Heat Advection/Diffusion& $\mathrm{Pe_{T}}$ && \textbf{8.1} (7.1--9.1)\e7 \\
Sensible/Latent Heat & $\mathrm{St}$ && \textbf{3.2}, $\infty$ \\
\end{tabular}
\caption{\textbf{Model parameters, dimensionless numbers, and their values}; reference values in bold; range of values explored in parentheses. (1) \cite{Moussallam2013}, (2) \cite{Klein2002}, (3) \cite{Oppenheimer2008}, (4) \cite{Molina2012}, (5) \cite{Giordano2008} , (6) \cite{Costa2009}.}
\label{tbl:RefParameters}
\end{center}
\end{table}

\begin{figure}
\begin{center}
\includegraphics[width=4in]{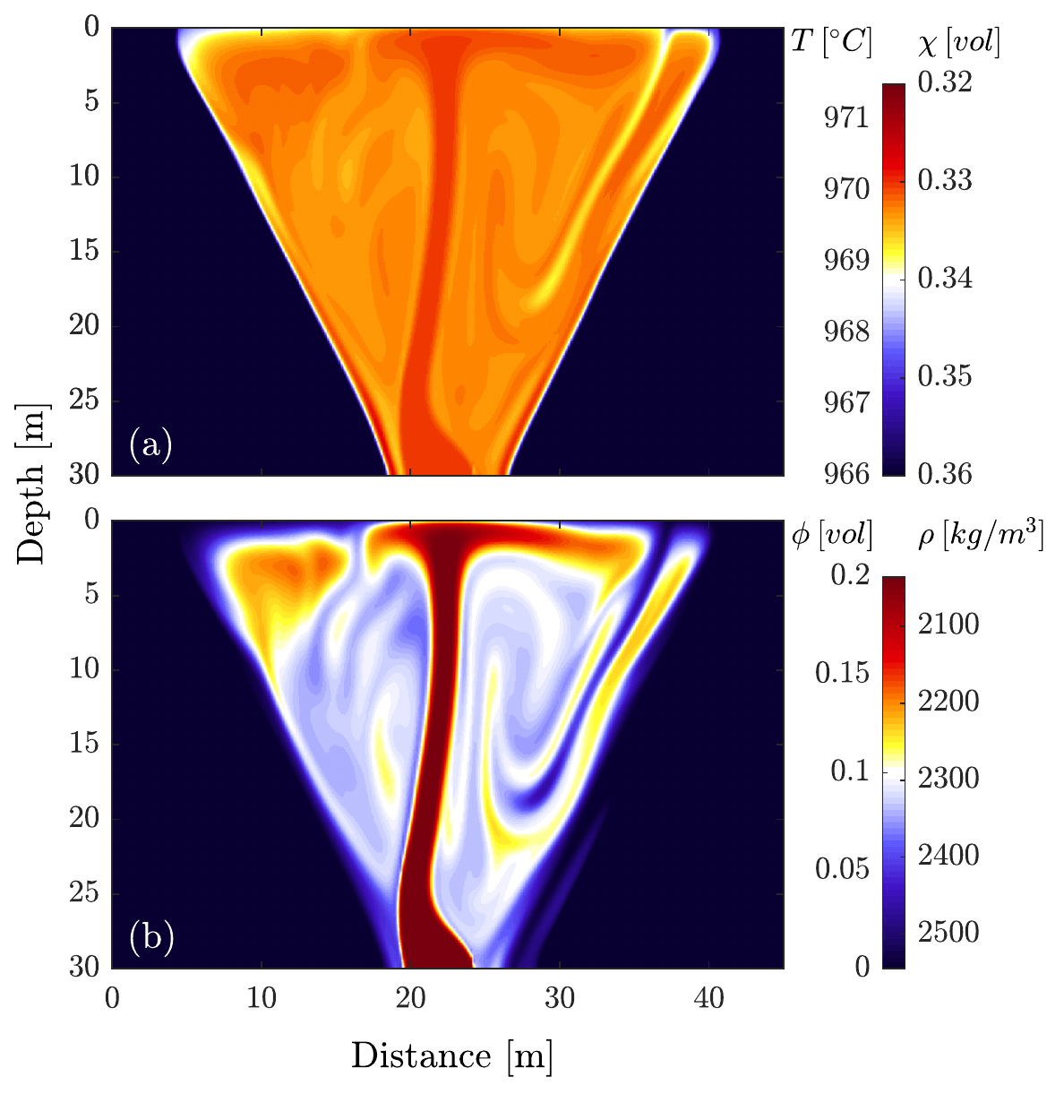}\\
\caption{(a) Example output of temperature and crystallinity field from t = 2 hrs in the reference parameter simulation which are equivalent by formulation, as a result of assuming near-equilibrium crystallization; (b) example density and vesicularity field. Density variation due to crystallinity in the lake interior is less than 5$\%$ of the total density variation such that vesicularity is a good approximation of the overall density.}
\label{fig:ref_panels}
\end{center}
\end{figure}

Figure \ref{fig:ref_panels} shows a snapshot of temperature, crystallinity, vesicularity, and density fields from the reference parameter test. The crystallinity directly follows the distribution of temperature and the two are therefore shown together in Fig.~\ref{fig:ref_panels}a. The range of crystallinity remains small within the lake. Hence, it imparts only minor variations to the density field, which is dominated by the effects of vesicularity (Fig.~\ref{fig:ref_panels}b). The density/vesicularity field shows the upwelling, gas-rich magma contrasting with the downwelling, outgassed magma. 
\par

\begin{figure}
\begin{center}
\includegraphics[width=\textwidth]{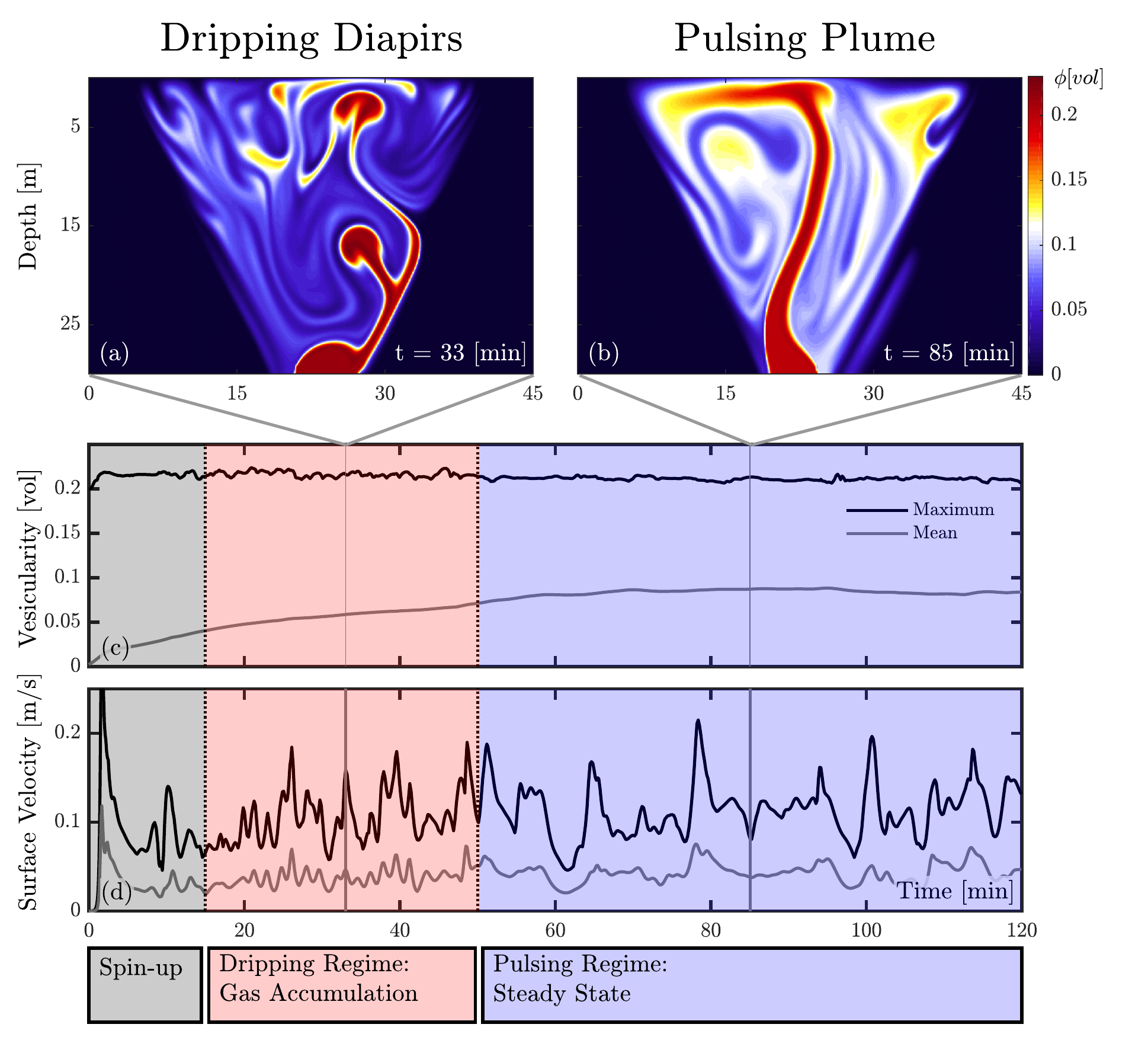}\\
\caption{Our reference simulation shows a transition between the dripping (a) and pulsing (b) regimes at $\sim$50 min. Panel c shows the vesicularity of the lake interior which increases away from the initial outgassed condition during the dripping regime before reaching a time-averaged steady state in the pulsing regime. Panel d shows the maximum and mean surface velocities of the lake surface. The difference in the surface velocity records of the two regimes should allow us to distinguish between regimes using surface observables.}
\label{fig:Ref_Regimes}
\end{center}
\end{figure}

Over the duration of the reference simulation, we observe two distinct flow regimes. The first 15 min of the simulation mark a spin-up period in which flow disrupts the initial conditions. The first persistent flow regime is characterized by a dripping instability in which roughly equant diapirs rise from the base of the lake or from upwelling magma along the lake wall (Fig.~\ref{fig:Ref_Regimes}a). We call this regime the \textit{dripping diapirs} or simply \textit{dripping} regime. After $\sim 50$ min, the flow switches to a new regime in which a near-continuous buoyant plume of magma ascends directly from the conduit mouth and rises through the center of the lake (Fig.~\ref{fig:Ref_Regimes}b). The plume oscillates from side to side and accommodates occasional pulses of gas-rich magma. We call this regime the \textit{pulsing plume} or simply \textit{pulsing} regime. The time required to reach the latter flow regime motivates the model run time of two hours applied to all parameter variations. \par

We suggest that the transition between dripping and pulsing regimes is the result of increased mean vesicularity in the lake. Gas-rich magma that reaches the surface is only partially outgassed before being recirculated, resulting in a gradual increase in the lake's gas content (Fig.~\ref{fig:Ref_Regimes}c). The transition between the regimes occurs when the mean vesicularity is $\sim0.07$, shortly before the vesicularity reaches a dynamic steady state of $0.085 \pm 0.005$ ($2 \sigma$). As the mean vesicularity increases, the density contrast between gas-rich upwelling and partially outgassed downwelling magmas decreases. The loss of relative buoyancy reduces the magma ascent rate compared to the conduit inflow and thus suppresses the dripping instability initially present at the conduit mouth. Visual inspection of the maximum and mean velocity time series (Fig.~\ref{fig:Ref_Regimes}d) further suggests that the different surface velocity patterns associated with the two regimes may be a useful metric to discriminate between the regimes in field data. \par

\subsection{Flow regimes}
The two flow regimes are controlled by the dimensionless numbers directly affecting the flux of gas into and out of the lake. These are the inflow number, $\mathrm{R_{in}}$ \eqref{eq:R_segr}, and the outgassing number, $\mathrm{Da_\phi}$ \eqref{eq:Da_phi}, which among other parameters are governed by the inflow vesicularity and rate, and the time scale of outgassing, respectively. We further investigate the physical mechanism underpinning the two regimes and their stability fields by testing the parameter variations given in Table \ref{tbl:ParameterVariations} whose results are visualized in the Supplemental material. \par

\begin{landscape}
\centering 
\begin{table}
\begin{center}
\begin{tabular}{l c c c c c}
Parameter & Symbol & Units & Trials & $\mathrm{R_{in}}$ (\e{-2}) & $\mathrm{Da_\phi}$ (\e{-3})\\
\hline\hline
Inflow Vesicularity & $\phi_0$ & Vol & 0.1, \textbf{0.2}, 0.3 &
9.9, \textbf{5.0}, 3.3 & 3.4, \textbf{3.9}, 4.4\\
Inflow Velocity & $u_{in}$ & m/s & 0.1, \textbf{0.2}, 0.3, 0.4, 0.5 &
2.5, \textbf{5.0}, 7.4, 9.9, 12 & \textbf{3.9} \\
Outgassing Time & $\tau_\phi$ & s & 40, 50, \textbf{80}, 160, $\infty$ &
\textbf{5.0} & 7.7, 6.2, \textbf{3.9}, 1.9, 0 \\
\hline
\end{tabular}
\caption{\textbf{Parameter variations for testing convective regimes}. Reference values highlighted in bold.}
\label{tbl:ParameterVariations}
\end{center}
\end{table}
\end{landscape}

\subsubsection{Dripping diapirs regime}
\begin{figure}
\begin{center}
\includegraphics[width=1.0\textwidth]{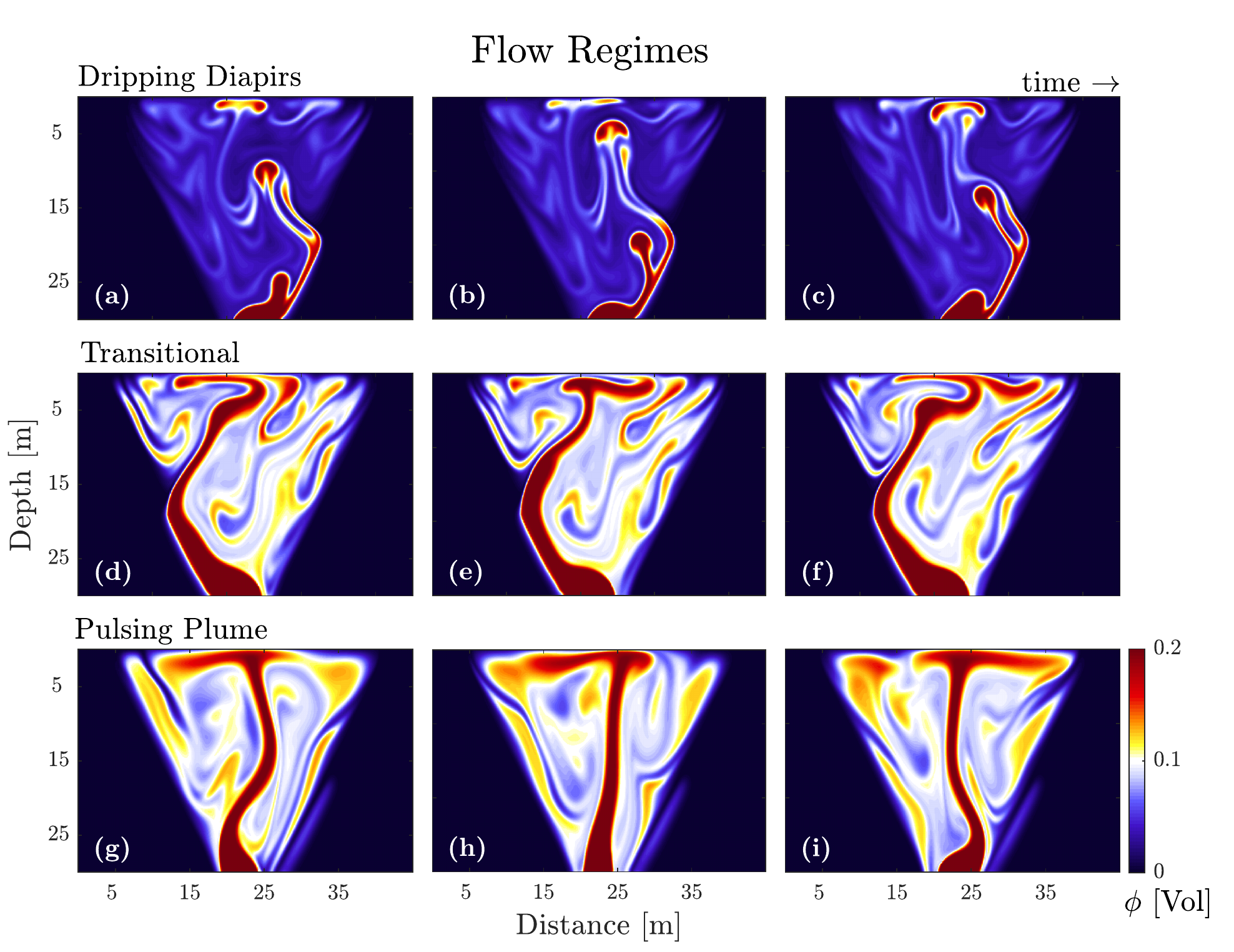}\\
\caption{Time evolution of the flow field in each regime. Panels a-c show the formation and ascent of diapirs in the dripping regime, d-f show pulses of magma travelling up an unsteady plume at the edge of the lake that characterize the transitional regime, and g-i show lateral migration of a free-standing plume in the pulsing regime.}
\label{fig:time_evolution}
\end{center}
\end{figure}

The dripping diapirs regime is characterized by an intermittent rise of buoyant, gas-rich magma diapirs formed by a dripping instability at the lake inlet. The diapirs either rise directly from the conduit or detach from magma that has ascended as far as one-half to two-thirds of the height of the lake along one of the walls (Fig.~\ref{fig:time_evolution}a-c). Diapirs have a radius of $\sim1/2$ the conduit radius and rise every $\sim1.5$--4 min at speeds of $\sim$ 0.2-0.4 m/s. \par

We find the dripping regime is stable when $\mathrm{Da_\phi}$ is high and $\mathrm{R_{in}}$ is low (Fig.~\ref{fig:Regime_Diagram}). $\mathrm{Da_\phi}$ is primarily controlled by the imposed surface outgassing time, $\tau_\phi$. When $\tau_\phi$ is short relative to time scale of convection, gas is removed efficiently from the system and a high density contrast between upwelling and outgassed magma is preserved. While this parameter is particular to our formulation and combines the effects of a range of processes, it highlights the importance of rapid outgassing in preventing gas build-up. \par

$\mathrm{R_{in}}$ depends linearly on the imposed inflow speed, which remains poorly constrained by field observations. We consider values between 0.1-0.5 m/s based on estimates from \citet{Calkins2008} and \citet{Oppenheimer2009}. Alternatively, an increase in inflow vesicularity will slightly reduce $\mathrm{R_{in}}$ since more buoyant magma will rise at a higher ascent speed relative to the inflow. Both slower inflow from the conduit as well as faster convective ascent speed promote stability of the dripping regime. \par

\subsubsection{Pulsing plume regime}
In the pulsing regime, buoyant magma rises through a mostly continuous plume of magma near the center of the lake (Fig.~\ref{fig:time_evolution}g-h). The plume does not steadily approach the surface, but oscillates from side to side with time. The plume is occasionally disrupted by diapirs that assemble at the conduit mouth or from the lateral extreme of a bend in the plume. At times, the downwelling flow cuts across the plume near the inlet. Even when continuous, the plume does not maintain a constant flux but accommodates pulses of increased flux ascending along the existing pathway. Pulses arise from a dripping instability similar to diapirs, but with a smaller volumetric imbalance. \par

\begin{figure}
\begin{center}
\includegraphics[width=\textwidth]{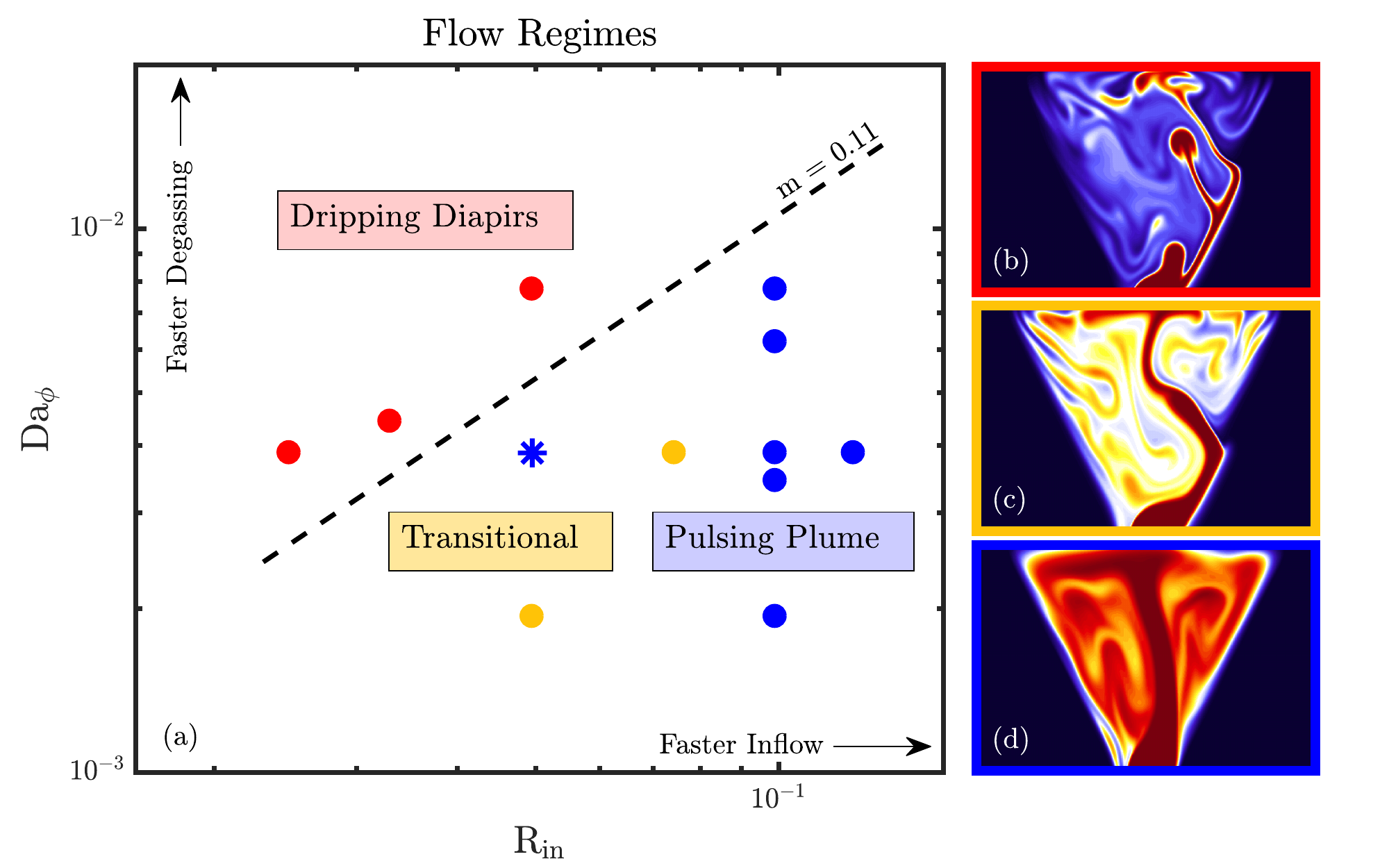}\\
\caption{Dimensionless parameters $\mathrm{R_{in}}$ and $\mathrm{Da_\phi}$ control the convective regime of the lava lake. Panel a suggests that the dripping regime is favored when outgassing is rapid and inflow is slow with respect to convective ascent and the pulsing regime dominates in the opposite case. The distinguishing flow patterns for the dripping (b), transitional (c), and pulsing (d) regimes highlight the ability of small changes in the physical parameters to alter the convective behavior.}
\label{fig:Regime_Diagram}
\end{center}
\end{figure}

We find the pulsing regime is stable when $\mathrm{Da_\phi}$ is small and $\mathrm{R_{in}}$ is large (Fig.~\ref{fig:Regime_Diagram}). For slow outgassing, the steady-state lake vesicularity is high, which reduces the ascent speed of gas-rich magma. The regime transition depends on gas in- and out-flux such that the pulsing regime is favored when $\mathrm{Da_\phi}$/$\mathrm{R_{in}} \lesssim 0.1$ (Fig.~\ref{fig:Regime_Diagram}). Within the range of outgassing rates we consider, conditions for the pulsing regime are met for $u_{in} \gtrsim$ 0.3 m/s and for $\phi_0 \gtrsim$ 0.2 We do not test scenarios of $\mathrm{R_{in}} \gtrsim 1$, where inflow exceeds free convection, which may be applicable for advection-dominated lava lakes like Ambrym \citep{Carniel2003,Lev2019}. 

\subsubsection{Transitional regime}
We observe a transitional regime between dripping and pulsing in which magma ascends as an unsteady plume along the lake walls in the bottom half of the lake before becoming detached and rising through the lake interior to the surface (Fig.~\ref{fig:time_evolution}g-i). The plume is more prone to disruption and diapirs from the conduit are more common in this than in the pulsing regime. Furthermore, the regime is transitional in that it sometimes marks a transition period between dripping and pulsing regimes as gas accumulates in the lake. In the reference case, this transition occurs quickly, in less than a single overturn cycle; however, in other cases it can be more protracted. The transitional behavior can persist until the end of the simulation time for conditions at intermediate $\mathrm{R_{in}}$ (5.0\e{-2} $\lesssim \mathrm{R_{in}} \lesssim$ 7.4\e{-2}) and low $\mathrm{Da_\phi}$ (Fig.~\ref{fig:Regime_Diagram}). In the cases where transitional behavior is observed after two hours model time, we find the average lake vesicularity continues to increase, suggesting that it may still represent an intermediate stage before the pulsing regime is reached at steady-state. \par

\subsubsection{Stagnant lid}
Since the focus of this paper is to understand the surface signatures of different convective regimes in an active lava lake, we are less interested in simulations leading to the formation of a stagnant lid. We observe the formation of a stagnant lid either if surface cooling is high or if outgassing is very inefficient. We find that to avoid a frozen stagnant lid, the cooling time must exceed $\sim$8 hr ($\mathrm{Da_T} \lesssim$ 1.1\e{-5}). Inefficient outgassing can lead to accumulation of gas-rich magma to the point that the vesicularity of the lake approaches that of the upwelling magma and gas-driven convection ceases (Supplementary Fig.~\ref{fig:Outgassing_Variations}e). We consider the limit where bubble segregation through the cooling skin is the only mechanism of gas removal ($\tau_\phi = \infty$) and find that process alone is too slow to maintain open convection. In that limit, other processes such as increased permeability through interconnected vesicularity or fissuring of the skin as it deforms must increase the rate of gas removal \citep{Blower2001}. \par

\section{Comparison to field data}
\subsection{Surface velocity}
\begin{figure}
\begin{center}
\includegraphics[width=\textwidth]{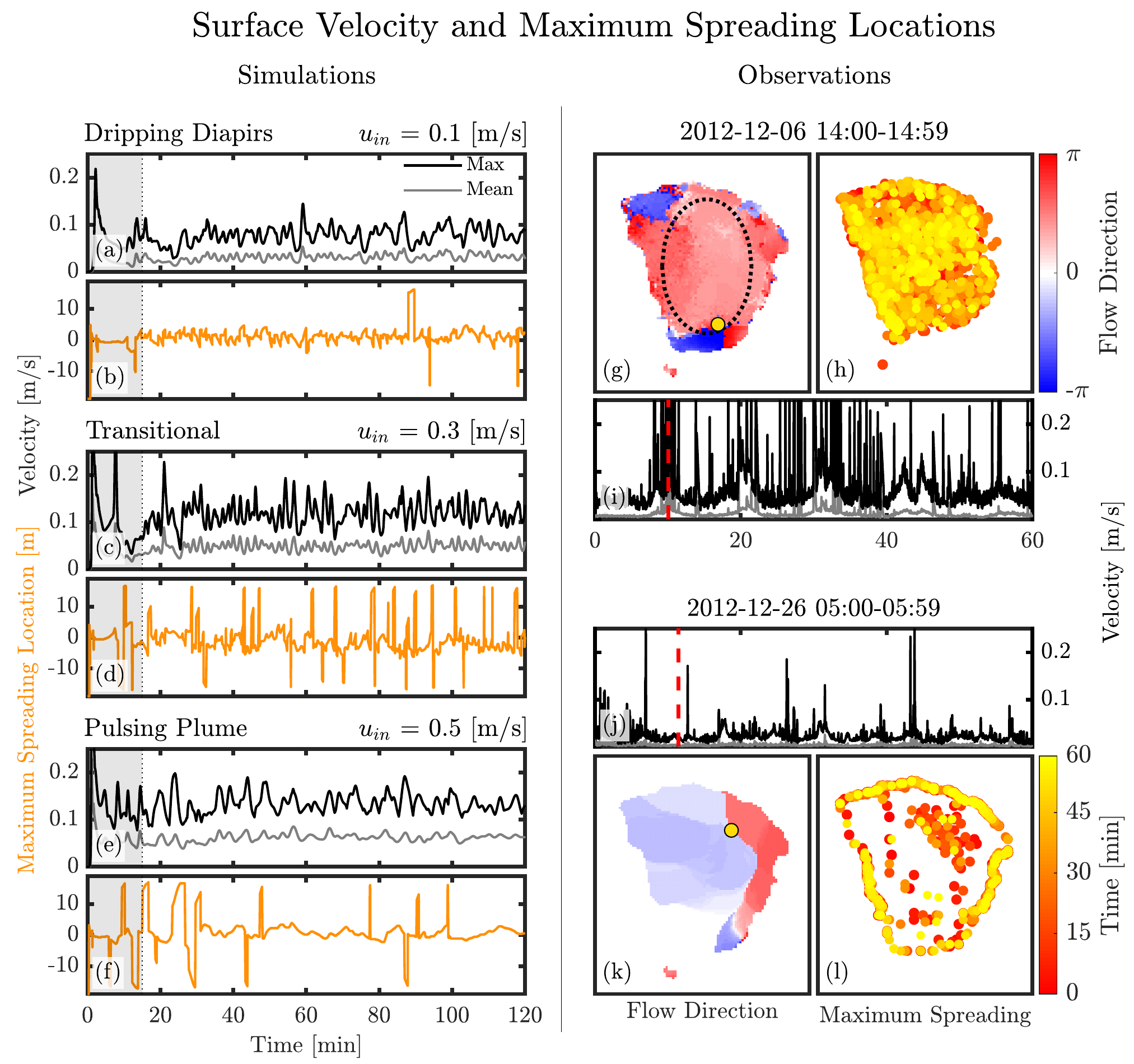}\\
\caption{Example surface velocity (a,c,e) and maximum spreading location (b,d,f) time series from model outputs demonstrate the surface characteristics of each regime. The dripping diapirs and transitional regimes show more rapid variation and spikier signals compared to the pulsing regime. Field observations also show considerable variability in surface velocity magnitude (i,j), direction (g,k), and locations of maximum divergence (h,l) which range from spikier signals with more varied locations of spreading and spreading directions (g-i) to intervals with more gradual changes in surface behavior (j-l). The elliptical mask used for the velocity magnitude Fourier analysis is shown in g, location of maximum divergence is found over the whole lake surface.}
\label{fig:Surface_Velocity_Data}
\end{center}
\end{figure}

We compare simulated surface data to field observations to evaluate which model regimes are applicable to the observed dynamics at Ray Lake. Fig.~\ref{fig:Surface_Velocity_Data}a,c $\&$ e) shows that the dripping and transitional regimes are characterized by more rapid variations in surface velocity than the pulsing regime. Surface velocity field data are calculated using optical velocimetry on thermal imaging observations from December 2012 (see \citet{Peters2014a} and \citet{Lev2019} for details on data collection and velocimetry). Here, we use two representative 1-hr long time series from 14:00-14:59 on Dec 06 and 05:00-05:59 on Dec 26 (Fig.~\ref{fig:Surface_Velocity_Data}i $\&$ j). These sequences were chosen for their low noise, lack of Strombolian eruptions, and to explore a range of behaviors exhibited by the lake. 
\par

We identify dominant periods of surface motion using the Fast Fourier Transform of the velocity in both the models and the field data. We exclude lake edges because they do not necessarily participate in the main convective pattern. For model output, we analyze the central 80$\%$ of the lake surface, and for observational data, use an elliptical mask centered on the lake (Fig.~\ref{fig:Surface_Velocity_Data}g). We normalize profiles at each node by their maximum value to reduce the bias toward the center of the lake that experiences higher mean velocities. The absolute values of model velocities range between 0.01 and 0.2 m/s, consistent with previous estimates \citep{Oppenheimer2009}. \par

\begin{figure}
\begin{center}
\includegraphics[width=\textwidth]{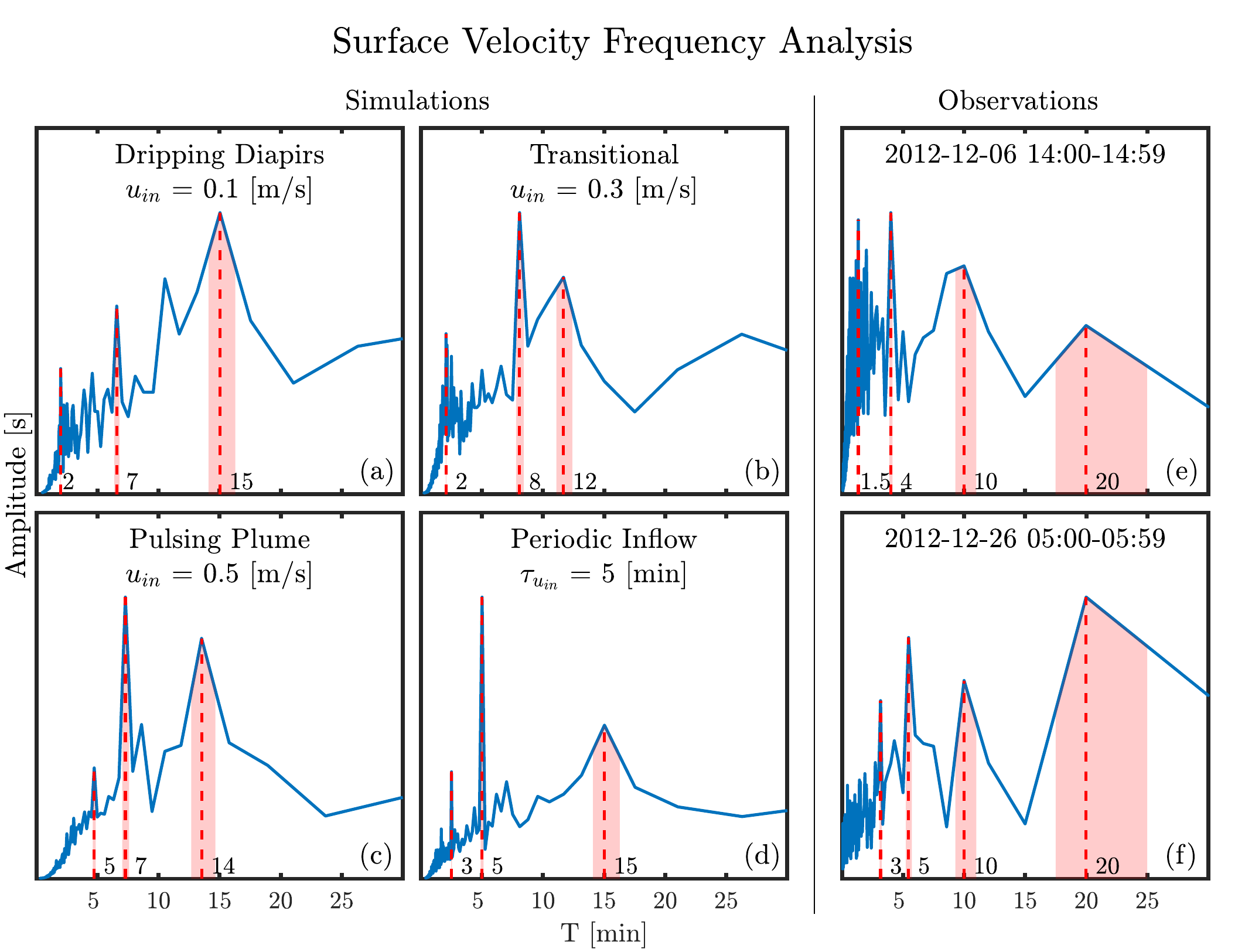}\\
\caption{Example velocity spectra from the dripping (a), transitional (b), and pulsing (c) regimes show several periodicities that reproduce periodicities found in the natural lake (e,f). Results from a simulation with an imposed conduit periodicity of 5 min show that the imposed periodicity is recoverable at the surface, but is modulated by other frequencies due to lake processes.}
\label{fig:Velocity_Spectra}
\end{center}
\end{figure}

The frequency analyses reveal several dominant periods in the simulations that are broadly consistent with those identified in the field data. Simulations in the dripping regime typically show dominant periods around 1.5-4 min, 7-9 min, and 11-15 min (Fig.~\ref{fig:Velocity_Spectra}a). The shortest period corresponds to the assembly and ascent of diapirs from the inlet, while the longer periods are related to lateral migration of the upwelling location and lake overturn. Simulations in the pulsing regime lack the shortest period found in the dripping regime but show peaks at 5-9 min, 12-15 min, and occasionally a weaker signal at 18-20 min associated with the lateral migration and disruption of the plume (see Fig.~\ref{fig:Velocity_Spectra}b, and Supplementary Fig.~\ref{fig:Velocity_Fourier_Supplement} for examples with 18-20 periods). Simulations in the transitional regime show periodicities similar to both the dripping and pulsing regimes. Our models predict a periodicity similar to that observed in the field data (Fig.~\ref{fig:Velocity_Spectra}e $\&$ f). However, we find that the analysis does not allow a clear distinction of the likely convective regime in Ray Lake. Short-period peaks (1.5--4 min) diagnostic of the dripping regime are  masked by noise in the field data and can therefore not be identified reliably enough to settle on a firm interpretation. \par

\subsection{Spreading location}
In addition to the velocity magnitudes, we also consider the spatial distribution of the divergence of surface velocity ($\Grad \cdot v$). At each model time step, we track the location of maximum divergence (spreading) at the surface, which we interpret as the upwelling location. \par 

The maximum spreading location time series are shown in Fig.~\ref{fig:Surface_Velocity_Data}b, d $\&$ f. The difference between the regimes is visually apparent: the dripping and transitional regimes are characterized by rapid variations, whereas the pulsing regime shows more continuous migration. We perform an equivalent two-dimensional analysis on the field data by tracking the maximum spreading location over time, in this case omitting the mask used in the velocity analysis to capture the full range of variability (Fig.~\ref{fig:Surface_Velocity_Data} h $\&$ l). The two observational records show distinctly different behavior. Data from Dec 6 shows greater variability in surface flow directions resulting in spreading locations covering the entire lake surface. The record from Dec 26 shows spreading locations concentrated at a few locations in the lake center or the lake edges arising a single spreading axis in the surface flow field. \par

\begin{figure}
\begin{center}
\includegraphics[width=\textwidth]{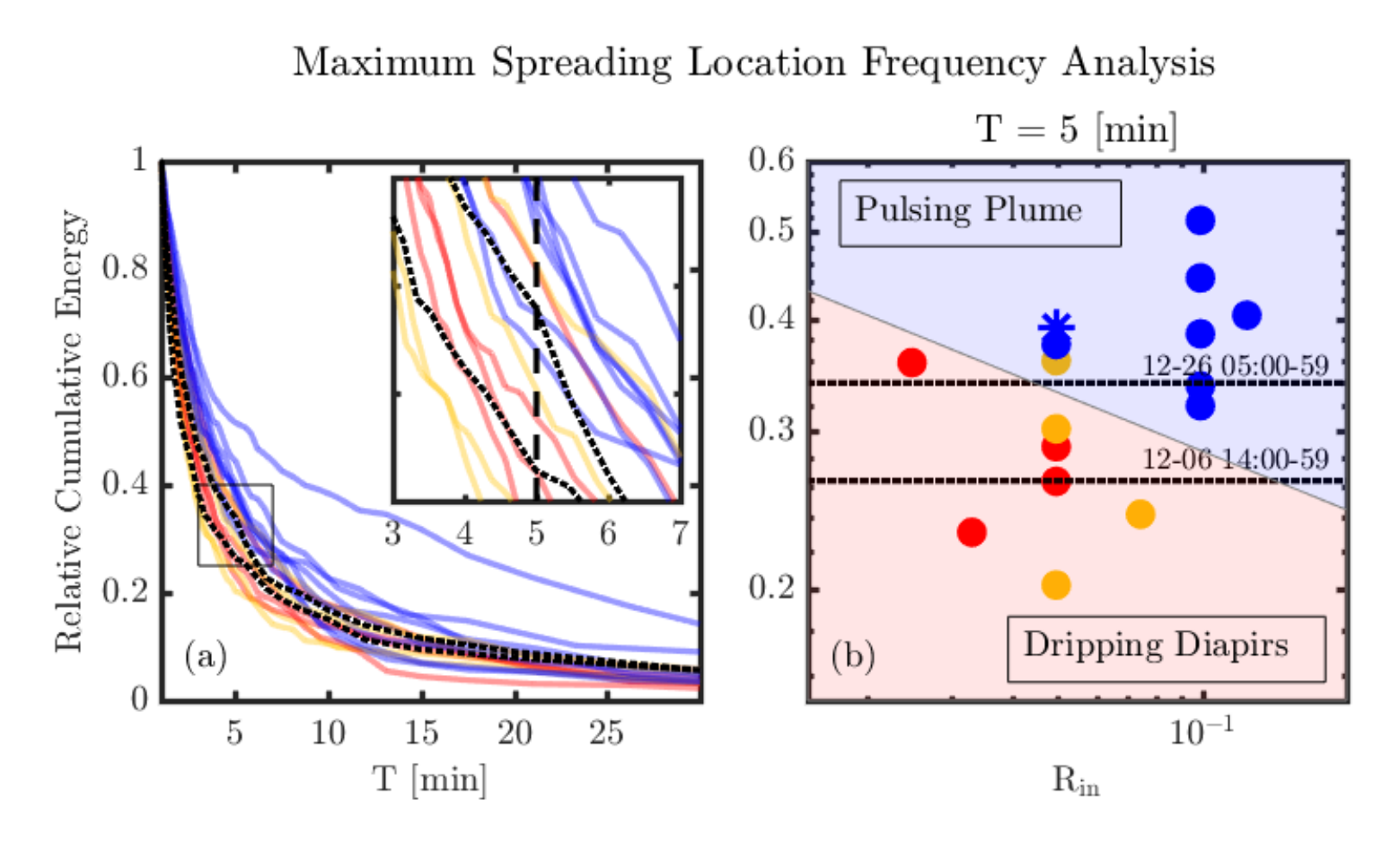}\\
\caption{The relative cumulative energy of the maximum spreading location period spectra (a) distinguishes between regimes, with more long-period ($T \geq 5$ min) energy found in the pulsing than the dripping regime (magnified view in inset panel). The cumulative energy at $T \geq 5$ min plotted against the inflow number $\mathrm{R_{in}}$ (b) reveals a clear regime boundary. The long-period energy taken from the two observational records overlaps the predicted boundary between dripping and pulsing regimes. In all panels the dripping regime shown in red, transitional in yellow, pulsing in blue, and field data in dashed black.}
\label{fig:Spreading_Location_Analysis}
\end{center}
\end{figure}

Similar to our analysis of surface velocities, we perform a Fourier analysis on time series of the maximum spreading location. The period spectra for field observations are quantified by averaging spectra of the distance from the lake center and of the azimuth relative to an arbitrary reference. Rather than seeking to identify specific short periods diagnostic of dripping, we consider the relative cumulative energy in longer periods ($T \geq 5$ min) of maximum spreading migration as shown in Fig.~\ref{fig:Spreading_Location_Analysis}. To reduce the effect of noise, we cut off energy below periods of 1 min, where we do not expect to find variability associated with the convective regime.  \par 

As expected based on the velocity magnitude analysis, we find that simulations in the pulsing regime have more energy above $T = 5$ min than those in the dripping regime. Within each regime, simulations with lower $\mathrm{R_{in}}$ (slower inflow, faster convection) have more long-period energy. The analysis of observational data returns values in the same range as the simulations, with the two observational records overlapping our predicted regime boundary. We thus interpret that dynamics at Ray Lake appear to straddle the regimes, and that the segment from Dec 6 shows behavior consistent with the dripping regime, whereas the segment from Dec 26 indicates the pulsing regime. Based upon analyses from other dates and times (not shown here), the Dec 26 data seem more representative of the typical behavior of the lake.\par

\subsection{Periodicities from the conduit}
Our analysis shows that steady inflow into a convecting lava lake can produce periodic behavior strikingly similar to observations. The question remains whether periodicities in conduit flow remain identifiable in surface observations. To identify how the lava lake modulates imposed conduit periodicities, we test inflow conditions with a sinusoidal time variation in inflow speed of a set period. We test inflow periodicity of 5, 10, and 20 min to cover the observed range and keep all other parameters the same as our reference case. Time-varying conduit inflow result in apparent switching between flow regimes over simulation time. When inflow is slow, dripping behavior is observed, and when inflow is fast, pulsing behavior. \par 

To determine whether the periodic inflow signal is recoverable at the surface, we analyze the surface velocity data in the same way as for constant inflow conditions. 
Figure \ref{fig:Velocity_Spectra}d shows the surface velocity spectrum recorded for a 5 min conduit inflow periodicity. The imposed periodicity from the conduit is indeed recoverable from the simulated surface velocity. However, the spectrum also shows the excitement of fundamental periods found in the reference simulation at constant inflow. For an imposed 5 min periodicity, we find spectral peaks at 3, 5, and 15 min, while the 7 min periodicity seen in the reference case is suppressed. We conclude that conduit periodicities can be expressed at the surface, but are overlaid with the fundamental periods pertaining to internal lake convection. A clear discrimination of conduit-related from lake-related signals remains challenging. \par 

\subsection{Model limitations}
Our choice of parameters is informed by field data, but we simplify a variety of processes including the removal of heat and gas from the lake, micro-physical phase interactions, boundary effects, and the potential role of gas slugs. In addition to these processes that are not currently represented in the model, the processes that are included might scale differently in the 2D model than in the 3D lake. Despite these simplifications, we find surprisingly good agreement between our model and field observations across a range of parameters. This may indicate that processes such as diapir formation and flow reorganization may be fundamental to lava lake circulation, and that the predicted behaviour is robust across a range of scales.  \par

With our reference value of $\tau_T$, simulations do not reach thermal equilibrium within the set run time of 2 hr. Rather, the mean lake temperature cools at approximately 0.65$\celsius$/hr. If this rate of heat loss continued linearly, we would expect the lake to fully crystallize after $>$120 hr. However, the majority of the heat lost is either by diffusion through the walls where thermal equilibrium cannot be achieved during the simulation time, or at the edges of the lake surface where the layer of magma is thin and rarely disrupted by convection. These top corners are the main source of thermal imbalance, and their contribution will decay with time as they reach thermal equilibrium (Supplementary Fig.~\ref{fig:Heat_Loss}). Accordingly, the convecting portion of the lake interior does reach an approximately steady-state temperature over the simulation time. \par

\section{Summary $\&$ conclusions}
Based on current observational and experimental constraints, our simulations suggest that the dynamics at Ray Lake may straddle the boundary between two convective regimes. In our model, shifts between the dripping and pulsing regimes are triggered by small changes in the relative speeds of conduit inflow to free convection ($\mathrm{R_{in}}$), and rates of surface outgassing to magma transport ($\mathrm{Da_\phi}$). Our results suggest that unsteady lava lake convection fed by a constant supply of gas-rich magma from the conduit is capable of generating periodic surface behavior similar to that observed at Ray Lake. We find that surface velocity spectra can be diagnostic of the convective regime, but that the most diagnostic short periods are typically masked by noise in the data. However, spectral energy in the migration of the maximum spreading location at periods longer than 5 min allows to distinguish more reliably between the dripping and pulsing regimes. Our findings further indicate that, using surface velocity data alone, periodic inflow from the conduit cannot be readily discriminated from unsteady lake convection. It is conceivable that observed variations in gas compositions may aid the interpretation of conduit conditions. We conclude that caution should be taken when interpreting surface records at lava lakes as a direct signature of conduit flow. \par

\section{Acknowledgements}
This research was supported by the Stanford Earth Summer Undergraduate Program and NSF under award EAR 1348022. T.K. acknowledges support from the Postdoc Mobility Fellowship 177816 by the Swiss National Science Foundation. E.L. was supported by NSF award 1348022.

\section{Author contributions}
This work is the result of the undergraduate thesis research of J.B., who performed the simulations and data analysis on model output and field data and took the lead in writing and figure preparation with the support of the other authors. T.K. and J.S. developed the research question, T.K. developed the numerical model, and E.L. provided the observational data for Mount Erebus, including image analysis and velocimetry. All authors reviewed and approved the text.

\section{Competing interests}
The authors have no competing interests to declare. 

\bibliography{ErebusReferences}

\begin{thebibliography}{}

\bibitem[\protect\astroncite{Arzi}{1978}]{Arzi1978}
Arzi, A.~A. (1978).
\newblock {Critical phenomena in the rheology of partially melted rocks}.
\newblock {\em Tectonophysics}, 44(1-4):173--184.

\bibitem[\protect\astroncite{Beckett et~al.}{2014}]{Beckett2014}
Beckett, F.~M., Burton, M., Mader, H.~M., Phillips, J.~C., Polacci, M., Rust,
  A.~C., and Witham, F. (2014).
\newblock {Conduit convection driving persistent degassing at basaltic
  volcanoes}.
\newblock {\em Journal of Volcanology and Geothermal Research}, 283:19--35.

\bibitem[\protect\astroncite{Beckett et~al.}{2011}]{Beckett2011}
Beckett, F.~M., Mader, H.~M., Phillips, J.~C., Rust, A.~C., and Witham, F.
  (2011).
\newblock {An experimental study of low-Reynolds-number exchange flow of two
  Newtonian fluids in a vertical pipe}.
\newblock {\em Journal of Fluid Mechanics}, 682:652--670.

\bibitem[\protect\astroncite{Blackburn et~al.}{1976}]{Blackburn1976}
Blackburn, E.~A., Wilson, L., and Sparks, R. S.~J. (1976).
\newblock {Mechanisms and dynamics of strombolian activity}.
\newblock {\em Journal of the Geological Society}, 132:429--440.

\bibitem[\protect\astroncite{Blower}{2001}]{Blower2001}
Blower, J.~D. (2001).
\newblock {Factors controlling permeability–porosity relationships in magma}.
\newblock {\em Bulletin of Volcanology}, 63(7):497--504.

\bibitem[\protect\astroncite{Brenner and Scott}{1994}]{Brenner1994}
Brenner, S.~C. and Scott, L.~R. (1994).
\newblock {\em {The Mathematical Theory of Finite Element Methods}}.
\newblock Springer, New York.

\bibitem[\protect\astroncite{Calkins et~al.}{2008}]{Calkins2008}
Calkins, J., Oppenheimer, C., and Kyle, P.~R. (2008).
\newblock {Ground-based thermal imaging of lava lakes at Erebus volcano,
  Antarctica}.
\newblock {\em Journal of Volcanology and Geothermal Research},
  177(3):695--704.

\bibitem[\protect\astroncite{Caricchi et~al.}{2007}]{Caricchi2007}
Caricchi, L., Burlini, L., Ulmer, P., Gerya, T., Vassalli, M., and Papale, P.
  (2007).
\newblock {Non-Newtonian rheology of crystal-bearing magmas and implications
  for magma ascent dynamics}.
\newblock {\em Earth and Planetary Science Letters}, 264:402--419.

\bibitem[\protect\astroncite{Carniel et~al.}{2003}]{Carniel2003}
Carniel, R., Di~Cecca, M., and Rouland, D. (2003).
\newblock {Ambrym, Vanuatu (July-August 2000): Spectral and dynamical
  transitions on the hours-to-days timescale}.
\newblock {\em Journal of Volcanology and Geothermal Research}, 128(1-3):1--13.

\bibitem[\protect\astroncite{Costa}{2005}]{Costa2005}
Costa, A. (2005).
\newblock {Viscosity of high crystal content melts: Dependence on solid
  fraction}.
\newblock {\em Geophysical Research Letters}, 32(22):1--5.

\bibitem[\protect\astroncite{Costa et~al.}{2009}]{Costa2009}
Costa, A., Caricchi, L., and Bagdassarov, N. (2009).
\newblock {A model for the rheology of particle-bearing suspensions and
  partially molten rocks}.
\newblock {\em Geochemistry, Geophysics, Geosystems}, 10(3).

\bibitem[\protect\astroncite{Csatho et~al.}{2008}]{Csatho2008}
Csatho, B., Schenk, T., Kyle, P., Wilson, T., and Krabill, W.~B. (2008).
\newblock {Airborne laser swath mapping of the summit of Erebus volcano,
  Antarctica: Applications to geological mapping of a volcano}.
\newblock {\em Journal of Volcanology and Geothermal Research},
  177(3):531--548.

\bibitem[\protect\astroncite{Dibble et~al.}{2008}]{Dibble2008}
Dibble, R.~R., Kyle, P.~R., and Rowe, C.~A. (2008).
\newblock {Video and seismic observations of Strombolian eruptions at Erebus
  volcano, Antarctica}.
\newblock {\em Journal of Volcanology and Geothermal Research}, 177:619--634.

\bibitem[\protect\astroncite{Drew}{1983}]{Drew1983}
Drew, D.~A. (1983).
\newblock {Mathematical Modeling of Two-Phase Flow}.
\newblock {\em Ann. Rev. Fluid Mech}, 15:261--91.

\bibitem[\protect\astroncite{Francis et~al.}{1993}]{Francis1993}
Francis, P., Oppenheimer, C., and Stevenson, D. (1993).
\newblock {Endogenous growth of persistently active volcanoes}.
\newblock {\em Letters to Nature}, 366:554--557.

\bibitem[\protect\astroncite{Fromm}{1968}]{Fromm1968}
Fromm, J.~E. (1968).
\newblock {A method for reducing dispersion in convective difference schemes}.
\newblock {\em Journal of Computational Physics}, 3:176--189.

\bibitem[\protect\astroncite{Giordano et~al.}{2008}]{Giordano2008}
Giordano, D., Russell, J.~K., and Dingwell, D.~B. (2008).
\newblock {Viscosity of magmatic liquids: A model}.
\newblock {\em Earth and Planetary Science Letters}, 271:123--134.

\bibitem[\protect\astroncite{Harris}{2008}]{Harris2008}
Harris, A.~J. (2008).
\newblock {Modeling lava lake heat loss, rheology, and convection}.
\newblock {\em Geophysical Research Letters}, 35(7):1--6.

\bibitem[\protect\astroncite{Harris et~al.}{2005}]{Harris2005}
Harris, A.~J., Carniel, R., and Jones, J. (2005).
\newblock {Identification of variable convective regimes at Erta Ale Lava
  Lake}.
\newblock {\em Journal of Volcanology and Geothermal Research},
  142(3-4):207--223.

\bibitem[\protect\astroncite{Heymann et~al.}{2002}]{Heymann2002}
Heymann, L., Peukert, S., and Aksel, N. (2002).
\newblock {On the solid-liquid transition of concentrated suspensions in
  transient shear flow}.
\newblock {\em Rheologica Acta}, 41(4):307--315.

\bibitem[\protect\astroncite{Huppert and Hallworth}{2007}]{Huppert2007}
Huppert, H. and Hallworth, M. (2007).
\newblock {Bi-directional flows in constrained systems}.
\newblock {\em J. Fluid Mech.}, 578:95--112.

\bibitem[\protect\astroncite{Kazahaya et~al.}{1994}]{Kazahaya1994}
Kazahaya, K., Shinohara, H., and Saito, G. (1994).
\newblock {Excessive degassing of Izu-Oshima volcano: magma convection in a
  conduit}.
\newblock {\em Bulletin of Volcanology}, 56:207--216.

\bibitem[\protect\astroncite{Keller et~al.}{2013}]{Keller2013}
Keller, T., May, D.~A., and Kaus, B. J.~P. (2013).
\newblock {Numerical modelling of magma dynamics coupled to tectonic
  deformation of lithosphere and crust}.
\newblock {\em Geophysical Journal International}, 195:1406--1442.

\bibitem[\protect\astroncite{Klein}{2002}]{Klein2002}
Klein, C. (2002).
\newblock {\em {The Manual of Mineral Science}}.
\newblock John Wiley {\&} Sons, Inc., 22 edition.

\bibitem[\protect\astroncite{Krieger and Dougherty}{1959}]{Krieger1959}
Krieger, I.~M. and Dougherty, T.~J. (1959).
\newblock {A Mechanism for Non-Newtonian Flow in Suspensions of Rigid Spheres}.
\newblock {\em Transactions of the Society of Rheology}, 137(1959).

\bibitem[\protect\astroncite{Le~Losq et~al.}{2015}]{LeLosq2015}
Le~Losq, C., Neuville, D.~R., Moretti, R., Kyle, P.~R., and Oppenheimer, C.
  (2015).
\newblock {Rheology of phonolitic magmas – the case of the Erebus lava lake}.
\newblock {\em Earth and Planetary Science Letters}, 411:53--61.

\bibitem[\protect\astroncite{Lev et~al.}{2019}]{Lev2019}
Lev, E., Spampinato, L., Patrick, M., Oppenheimer, C., Peters, N., Hernandez,
  P., and Marlow, J. (2019).
\newblock {A global sythesis of lava lake dynamics}.
\newblock {\em Journal of Volcanology and Geothermal Research}.

\bibitem[\protect\astroncite{Mader et~al.}{2013}]{Mader2013}
Mader, H.~M., Llewellin, E.~W., and Mueller, S.~P. (2013).
\newblock {The rheology of two-phase magmas: A review and analysis}.
\newblock {\em Journal of Volcanology and Geothermal Research}, 257:135--158.

\bibitem[\protect\astroncite{Manga}{1996}]{Manga1996}
Manga, M. (1996).
\newblock {Waves of bubbles in basaltic magmas and lavas}.
\newblock {\em Journal of Geophysical Research}, 101(B8):457--17.

\bibitem[\protect\astroncite{Molina et~al.}{2012}]{Molina2012}
Molina, I., Burgisser, A., and Oppenheimer, C. (2012).
\newblock {Numerical simulations of convection in crystal-bearing magmas: A
  case study of the magmatic system at Erebus, Antarctica}.
\newblock {\em Journal of Geophysical Research}, 117(April).

\bibitem[\protect\astroncite{Moussallam et~al.}{2013}]{Moussallam2013}
Moussallam, Y., Oppenheimer, C., Scaillet, B., Kyle, P.~R., Des, I., La, S.
  D.~E., and Orle, T.~D. (2013).
\newblock {Experimental Phase-equilibrium Constraints on the Phonolite Magmatic
  System of Erebus Volcano, Antarctica}.
\newblock {\em Journal of Petrology}, 54(7):1285--1307.

\bibitem[\protect\astroncite{Mucha et~al.}{2004}]{Mucha2004}
Mucha, P.~J., Tee, S.-Y., Weitz, D.~A., Shraiman, B.~I., and Brenner, M.~P.
  (2004).
\newblock {A model for velocity fluctuations in sedimentation}.
\newblock {\em Journal of Fluid Mechanics}, 501:71--104.

\bibitem[\protect\astroncite{Oppenheimer and Kyle}{2008}]{Oppenheimer2008}
Oppenheimer, C. and Kyle, P.~R. (2008).
\newblock {Probing the magma plumbing of Erebus volcano, Antarctica, by
  open-path FTIR spectroscopy of gas emissions}.
\newblock {\em Journal of Volcanology and Geothermal Research},
  177(3):743--754.

\bibitem[\protect\astroncite{Oppenheimer et~al.}{2009}]{Oppenheimer2009}
Oppenheimer, C., Lomakina, A.~S., Kyle, P.~R., Kingsbury, N.~G., and Boichu, M.
  (2009).
\newblock {Pulsatory magma supply to a phonolite lava lake}.
\newblock {\em Earth and Planetary Science Letters}, 284(3-4):392--398.

\bibitem[\protect\astroncite{Palma et~al.}{2011}]{Palma2011}
Palma, J.~L., Blake, S., and Calder, E.~S. (2011).
\newblock {Constraints on the rates of degassing and convection in basaltic
  open-vent volcanoes}.
\newblock {\em Geochemistry, Geophysics, Geosystems}, 12(11).

\bibitem[\protect\astroncite{Patrick et~al.}{2016}]{Patrick2016}
Patrick, M.~R., Orr, T., Sutton, A.~J., Lev, E., Thelen, W., and Fee, D.
  (2016).
\newblock {Shallowly driven fluctuations in lava lake outgassing (gas
  pistoning), Kilauea Volcano}.
\newblock {\em Earth and Planetary Science Letters}, 433:326--338.

\bibitem[\protect\astroncite{Peters et~al.}{2014a}]{Peters2014}
Peters, N., Oppenheimer, C., Killingsworth, D.~R., Frechette, J., and Kyle,
  P.~R. (2014a).
\newblock {Correlation of cycles in Lava Lake motion and degassing at Erebus
  Volcano, Antarctica}.
\newblock {\em Geochemistry, Geophysics, Geosystems}, 15(8):3244--3257.

\bibitem[\protect\astroncite{Peters et~al.}{2014b}]{Peters2014a}
Peters, N., Oppenheimer, C., Kyle, P., and Kingsbury, N. (2014b).
\newblock {Decadal persistence of cycles in lava lake motion at Erebus volcano,
  Antarctica}.
\newblock {\em Earth and Planetary Science Letters}, 395:1--12.

\bibitem[\protect\astroncite{Pistone et~al.}{2012}]{Pistone2012}
Pistone, M., Caricchi, L., Ulmer, P., Burlini, L., Ardia, P., Reusser, E.,
  Marone, F., and Arbaret, L. (2012).
\newblock {Deformation experiments of bubble- and crystal-bearing magmas:
  Rheological and microstructural analysis}.
\newblock {\em Journal of Geophysical Research}, 117.

\bibitem[\protect\astroncite{Renner et~al.}{2000}]{Renner2000}
Renner, Y., Evans, B., and Hirth, G. (2000).
\newblock {On the rheologically critical melt fraction}.
\newblock {\em Earth and Planetary Science Letters}, 181:585--594.

\bibitem[\protect\astroncite{Richardson and Zaki}{1954}]{Richardson1954}
Richardson, J.~F. and Zaki, W.~N. (1954).
\newblock {The sedimentation of a suspension of uniform spheres under
  conditions of viscous flow}.
\newblock {\em Chemical Engineering Science}, 3(2):65--73.

\bibitem[\protect\astroncite{Saar et~al.}{2001}]{Saar2001}
Saar, M.~O., Manga, M., Cashman, K.~V., and Fremouw, S. (2001).
\newblock {Numerical models of the onset of yield strength in crystal-melt
  suspensions}.
\newblock {\em Earth and Planetary Science Letters}, 187:367--379.

\bibitem[\protect\astroncite{Segre et~al.}{2001}]{Segre2001}
Segre, P.~N., Liu, F., Umbanhowar, P., and Weitz, D.~A. (2001).
\newblock {An effective gravitational temperature for sedimentation}.
\newblock {\em Nature}, 409(February):594--597.

\bibitem[\protect\astroncite{Stevenson and Blake}{1998}]{Stevenson1998}
Stevenson, D.~S. and Blake, S. (1998).
\newblock {Modelling the dynamics and thermodynamics of volcanic degassing}.
\newblock {\em Bulletin of Volcanology}, 60:307--317.

\bibitem[\protect\astroncite{Suckale et~al.}{2018}]{Suckale2018}
Suckale, J., Qin, Z., Picchi, D., Keller, T., and Battiato, I. (2018).
\newblock {Bistability of buoyancy-driven exchange flows in vertical tubes}.
\newblock {\em Journal of Fluid Mechanics}, 850:525--550.

\bibitem[\protect\astroncite{Sweeney et~al.}{2008}]{Sweeney2008}
Sweeney, D., Kyle, P.~R., and Oppenheimer, C. (2008).
\newblock {Sulfur dioxide emissions and degassing behavior of Erebus volcano,
  Antarctica}.
\newblock {\em Journal of Volcanology and Geothermal Research},
  177(3):725--733.

\bibitem[\protect\astroncite{Wright and Pilger}{2008}]{Wright2008}
Wright, R. and Pilger, E. (2008).
\newblock {Satellite observations reveal little inter-annual variability in the
  radiant flux from the Mount Erebus lava lake}.
\newblock {\em Journal of Volcanology and Geothermal Research}, 177:687--694.

\end{thebibliography}

\cleardoublepage 

\section*{Supplementary material}
\renewcommand{\thefigure}{S\arabic{figure}}
\setcounter{figure}{0}
\renewcommand{\thetable}{S\arabic{table}}
\setcounter{table}{0}

\begin{figure}[hbp]
\begin{center}
\includegraphics[width=100mm]{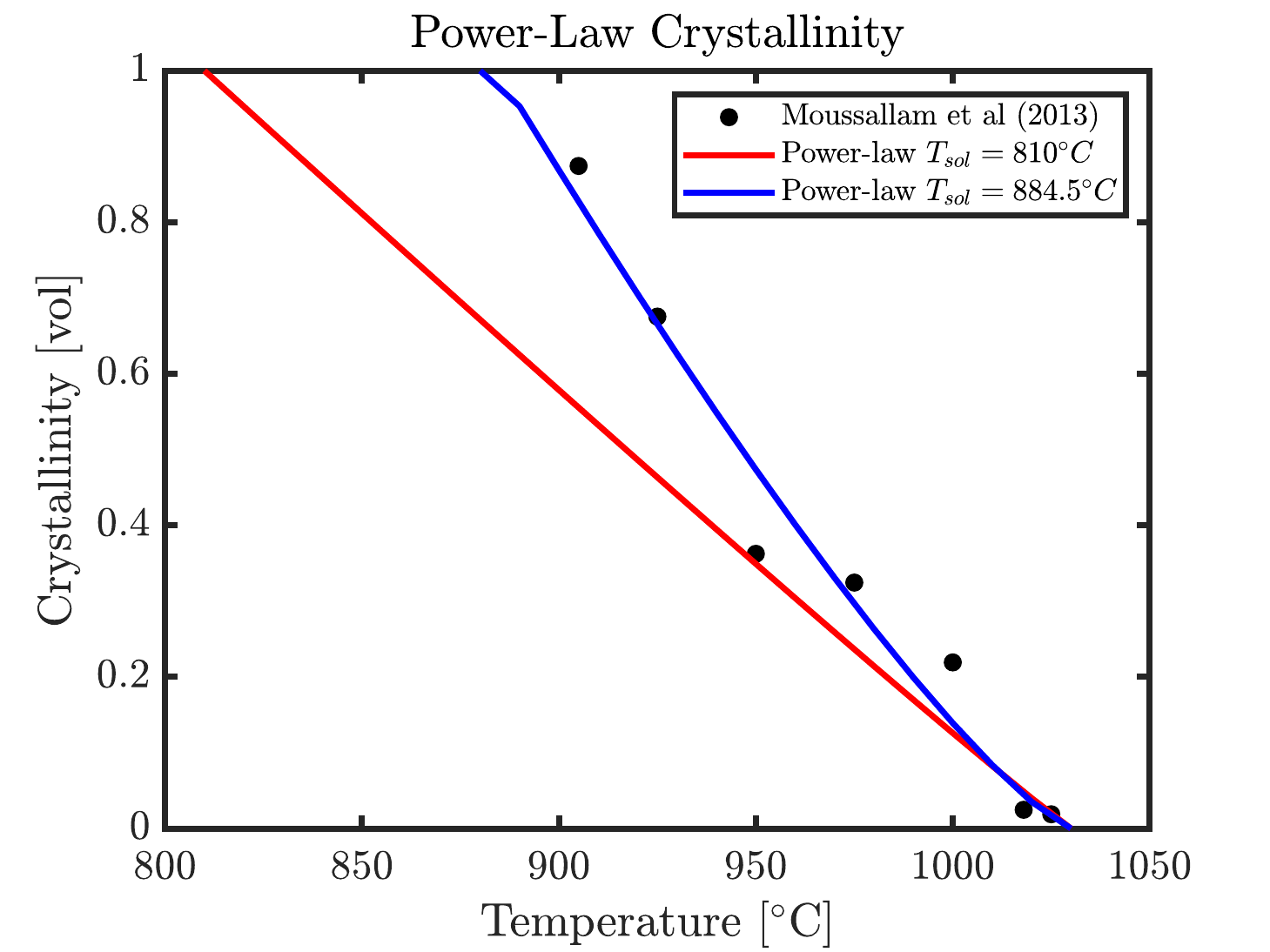}\\
\caption{Power-law fitting in temperature to equilibrium crystallinity data from \citet{Moussallam2013}). The power-law fit matches the data poorly when using the theoretical values for the solidus temperature (red). A fitted solidus temperature of 884.5$\celsius$ yields the best fit with the power-law crystallinity parameterization (blue).}
\label{fig:Crystallinity}
\end{center}
\end{figure}

\begin{figure}
\begin{center}
\includegraphics[width=100mm]{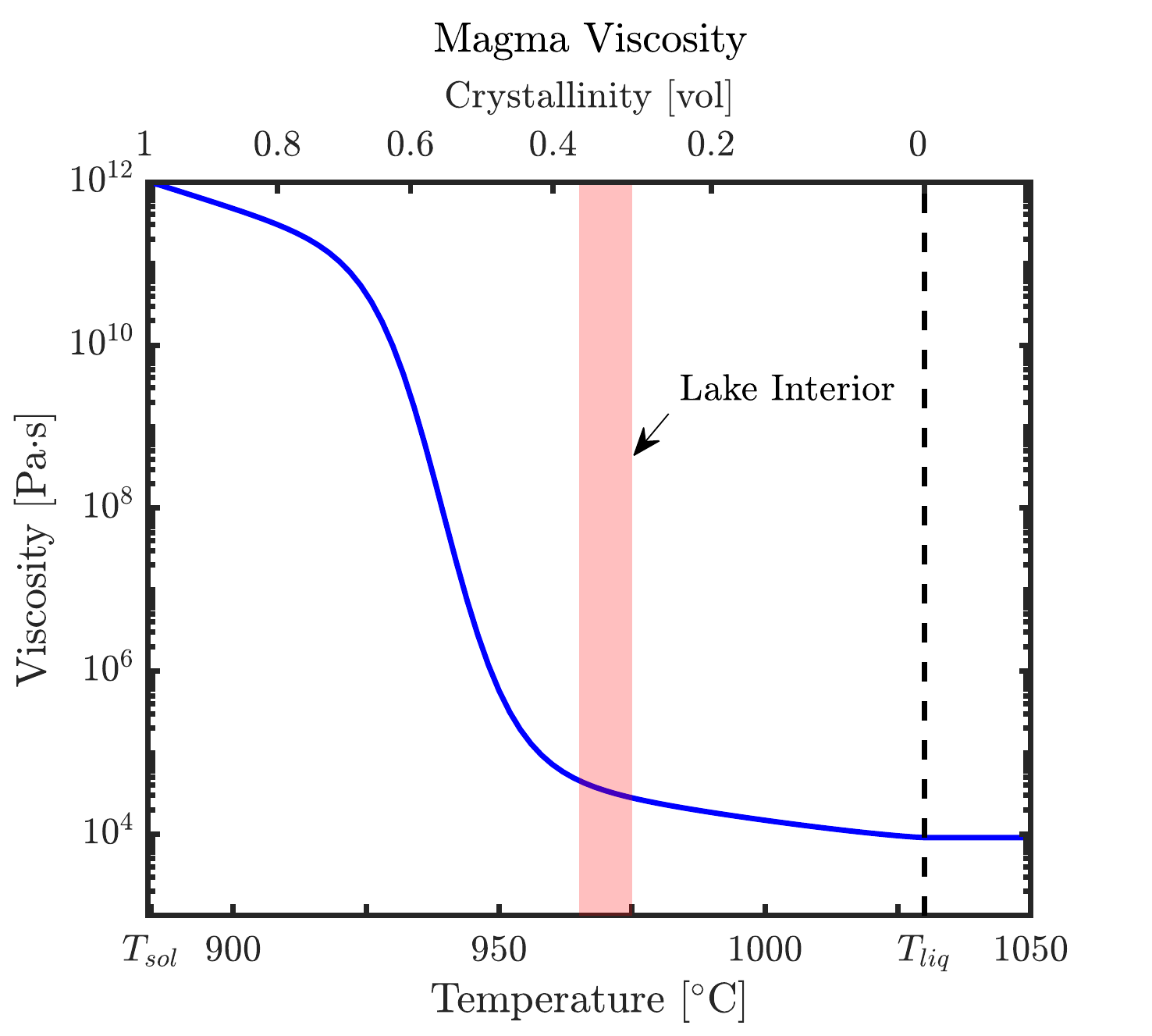}\\
\caption{Crystallinity and temperature dependence of viscosity at the reference strain rate of $10^{-3}$ s$^{-1}$. At low temperatures the viscosity approaches 10$^12$ Pa.s and at high temperatures the viscosity approaches the melt viscosity of 10$^4$ Pa.s.}
\label{fig:Viscosity}
\end{center}
\end{figure}

\begin{figure}
\begin{center}
\includegraphics[width=150mm]{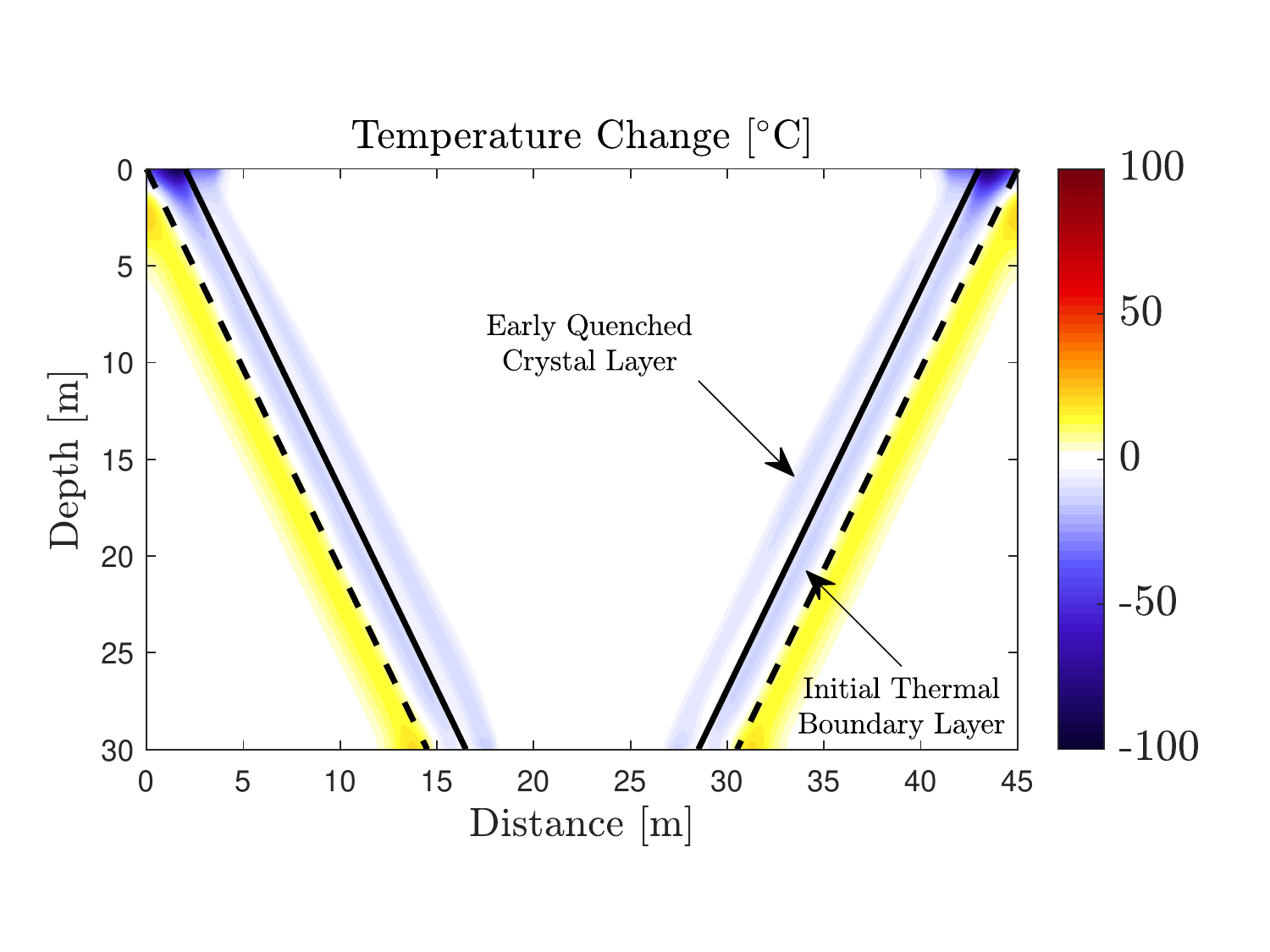}\\
\caption{Heat loss after two hours model time for the reference simulation. The convective lake interior maintains a near-constant temperature. Heat loss is primarily concentrated at the lake walls ($\Delta T < 15 \celsius$) and in the upper corners where surface heat loss dominates and convection does not efficiently recycle material.}
\label{fig:Heat_Loss}
\end{center}
\end{figure}

\begin{figure}
\begin{center}
\includegraphics[width=\textwidth]{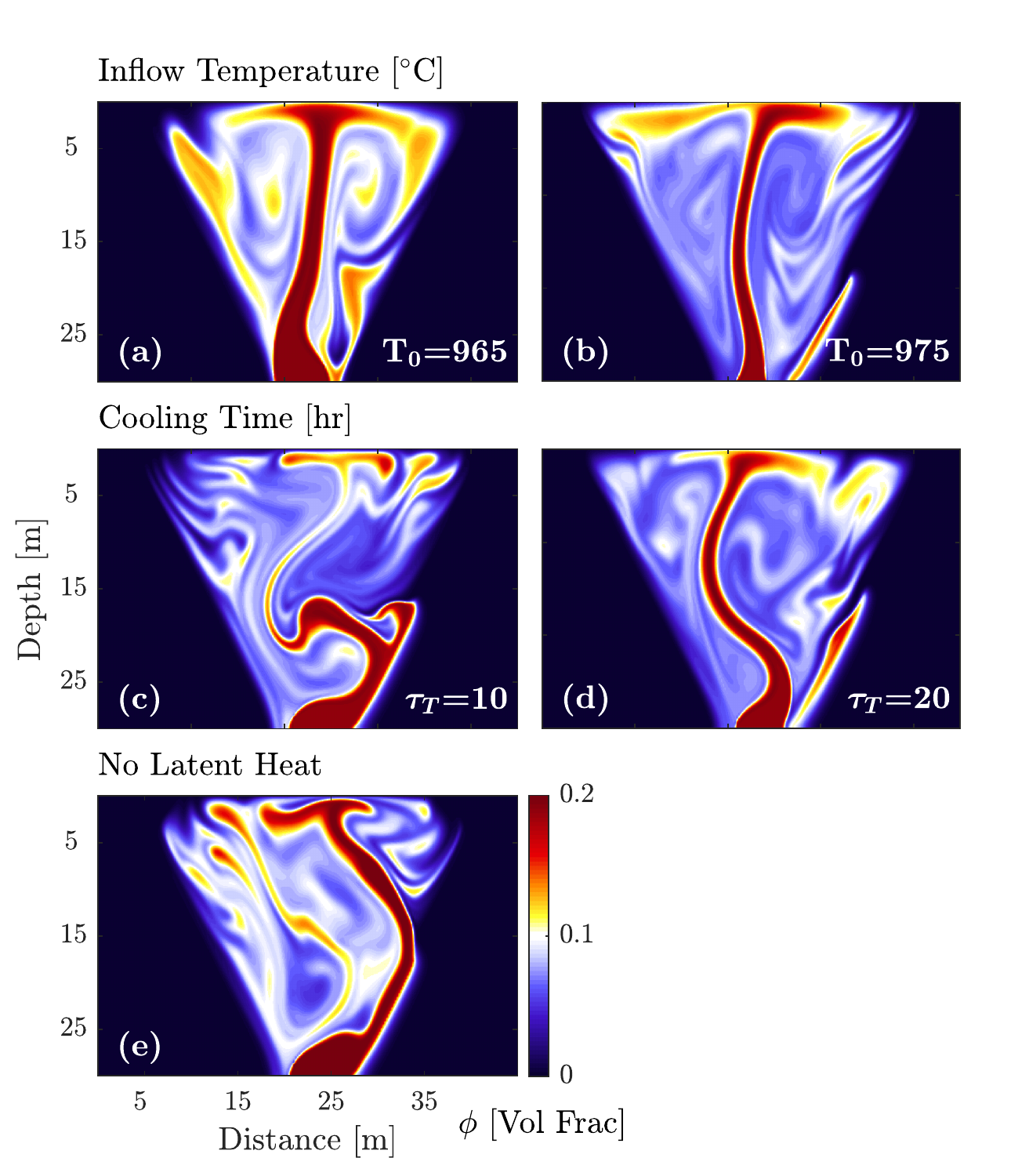}\\
\caption{Vesicularity after 2 hrs model time, showing variations in heat removal: changes in upwelling magma temperature (a, b), surface cooling time (c, d), and without latent heat (e). The main effect of temperature in the lake is through the stiffening effect of crystallinity on viscosity, but without variation in conduit forcing or gas removal, small changes in the temperature does not move simulations out of the transitional or pulsing plume regimes.}
\label{fig:Cooling_Variations}
\end{center}
\end{figure}

\begin{figure}
\begin{center}
\includegraphics[width=\textwidth]{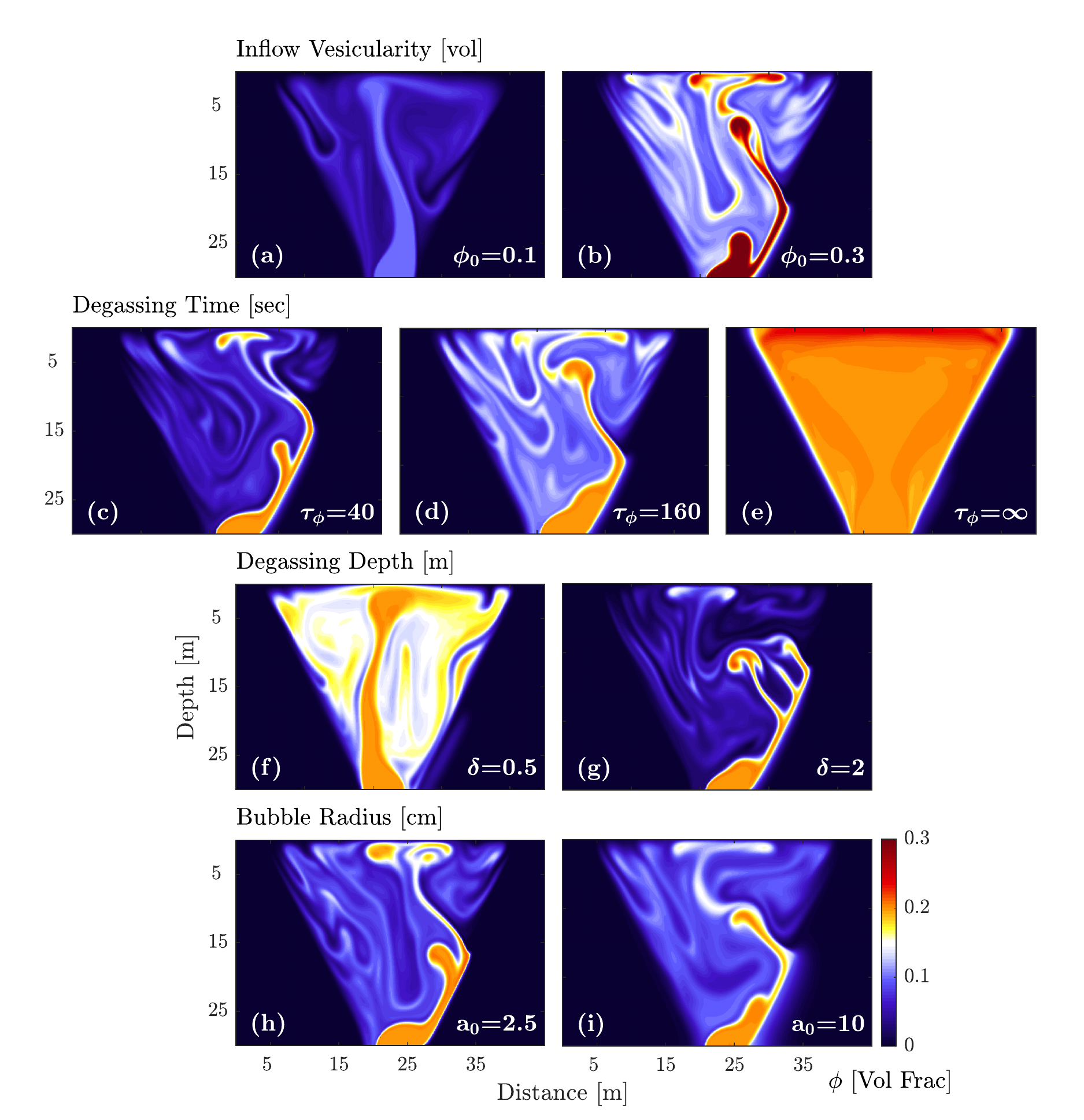}\\
\caption{Vesicularity after 2 hrs model time, showing variations in gas removal: changes in upwelling magma vesicularity (a, b), surface outgassing time (c, d, e), surface outgassing and cooling depth (f, g) that demonstrate the critical role of inflow vesicularity and outgassing rate on maintaining buoyancy contrast between up- and downwelling lava. Bubble radius within the range we explore does not have a strong effect on convection (h, i).}
\label{fig:Outgassing_Variations}
\end{center}
\end{figure}

\begin{figure}
\begin{center}
\includegraphics[width=\textwidth]{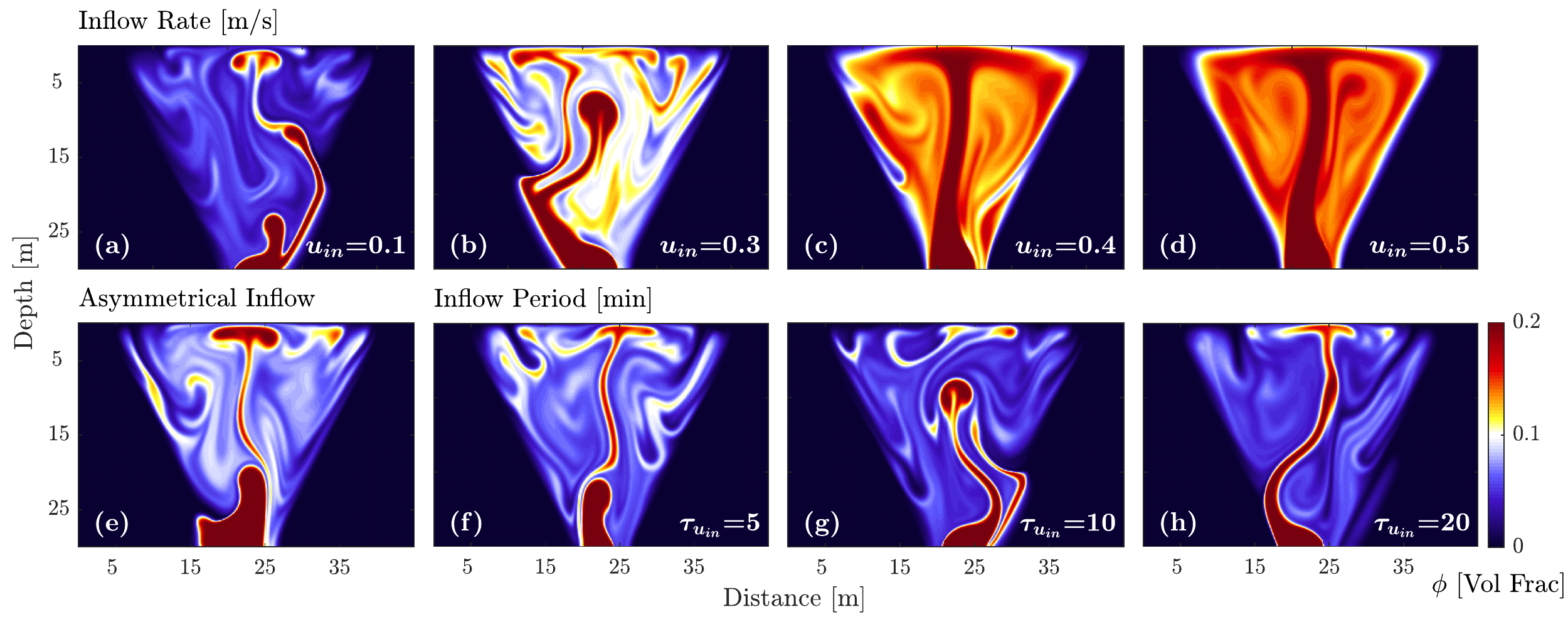}\\
\caption{Vesicularity after 2 hrs model time, showing variations inflow conditions. Changes in upwelling magma velocity (a-d) exert a strong control on flow behavior. Because our model imposes a velocity profile at the conduit, asymmetric inflow results in a large amount of forced downwelling of vesicular lava (e). Periodic inflow (f-h) modifies behavior between the regimes identified for constant flow, consistent with their instantaneous inflow velocity.}
\label{fig:Inflow_Variations}
\end{center}
\end{figure}

\begin{figure}
\begin{center}
\includegraphics[width=\textwidth]{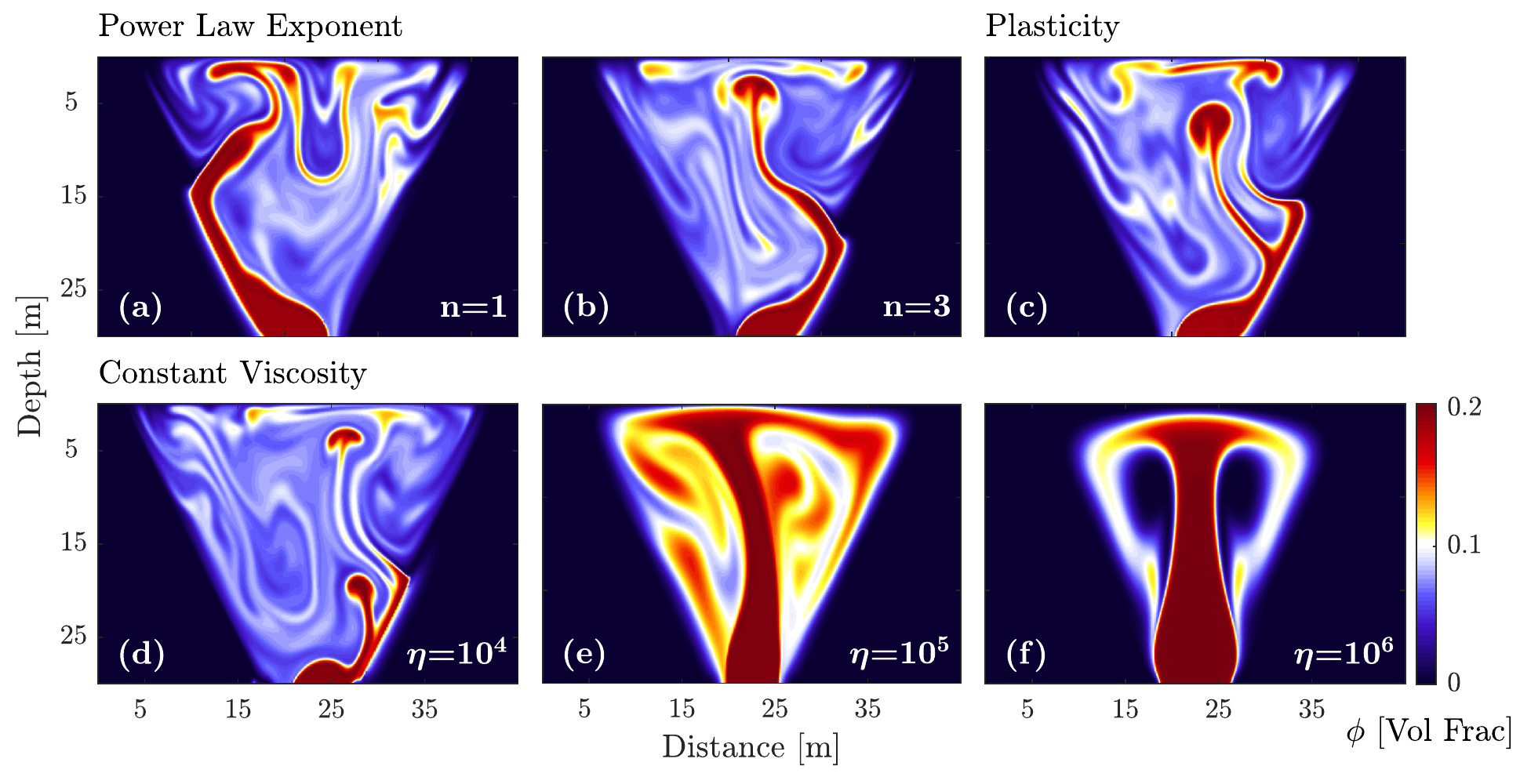}\\
\caption{Vesicularity after 2 hrs model time, showing variations magma rheology: changes in power-law exponent  (a, b), with Bingham yield strength (c), and constant viscosity (d, e, f). Viscosity is the controlling rheological parameter and very high viscosity results in stabilization of the upwelling plume.}
\label{fig:Rheology_Variations}
\end{center}
\end{figure}

\begin{figure}
\begin{center}
\includegraphics[width=\textwidth]{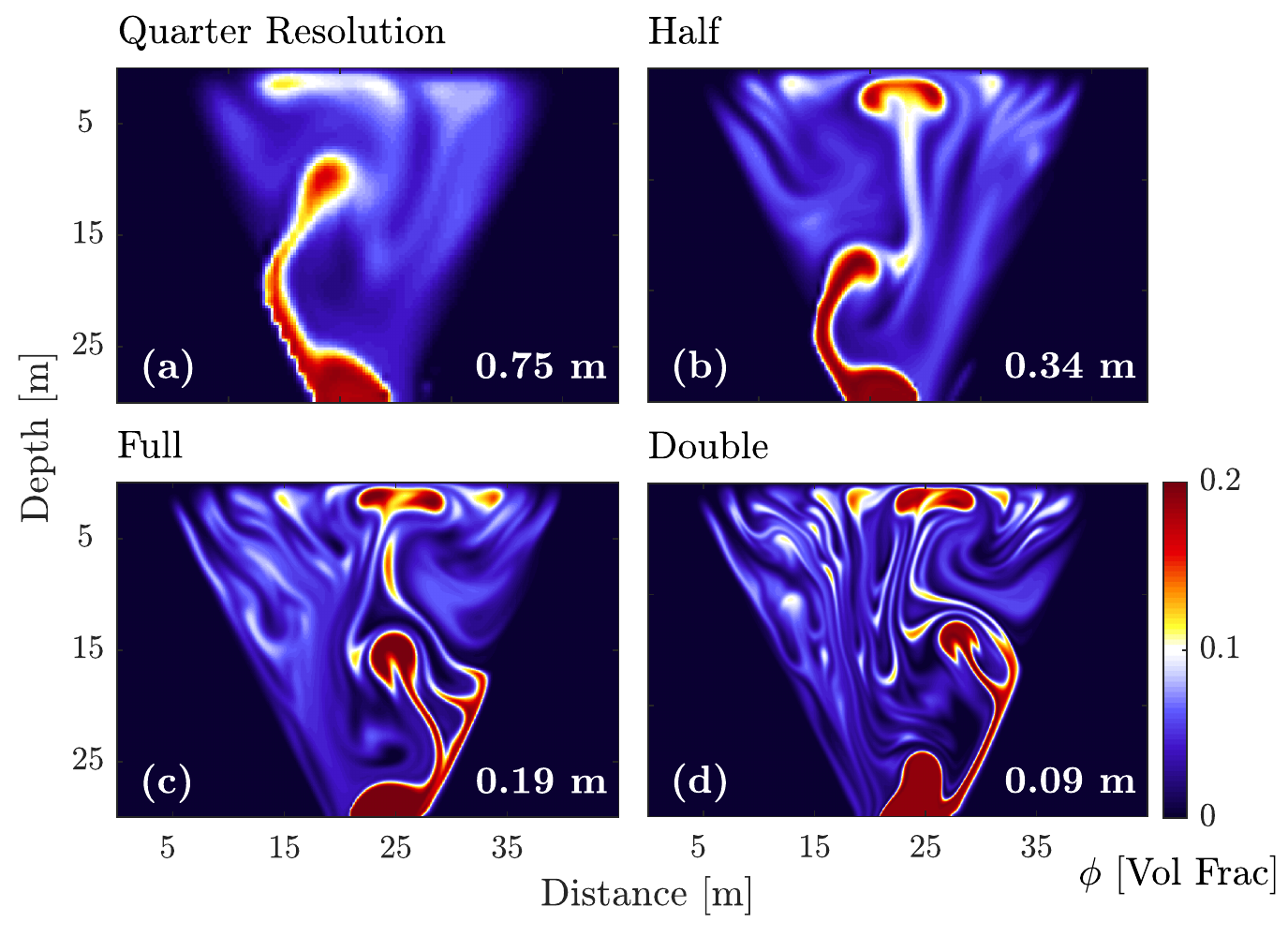}\\
\caption{Vesicularity after 20 min model time, showing convergence.}
\label{fig:Resolution_testing}
\end{center}
\end{figure}

\begin{figure}
\begin{center}
\includegraphics[width=\textwidth]{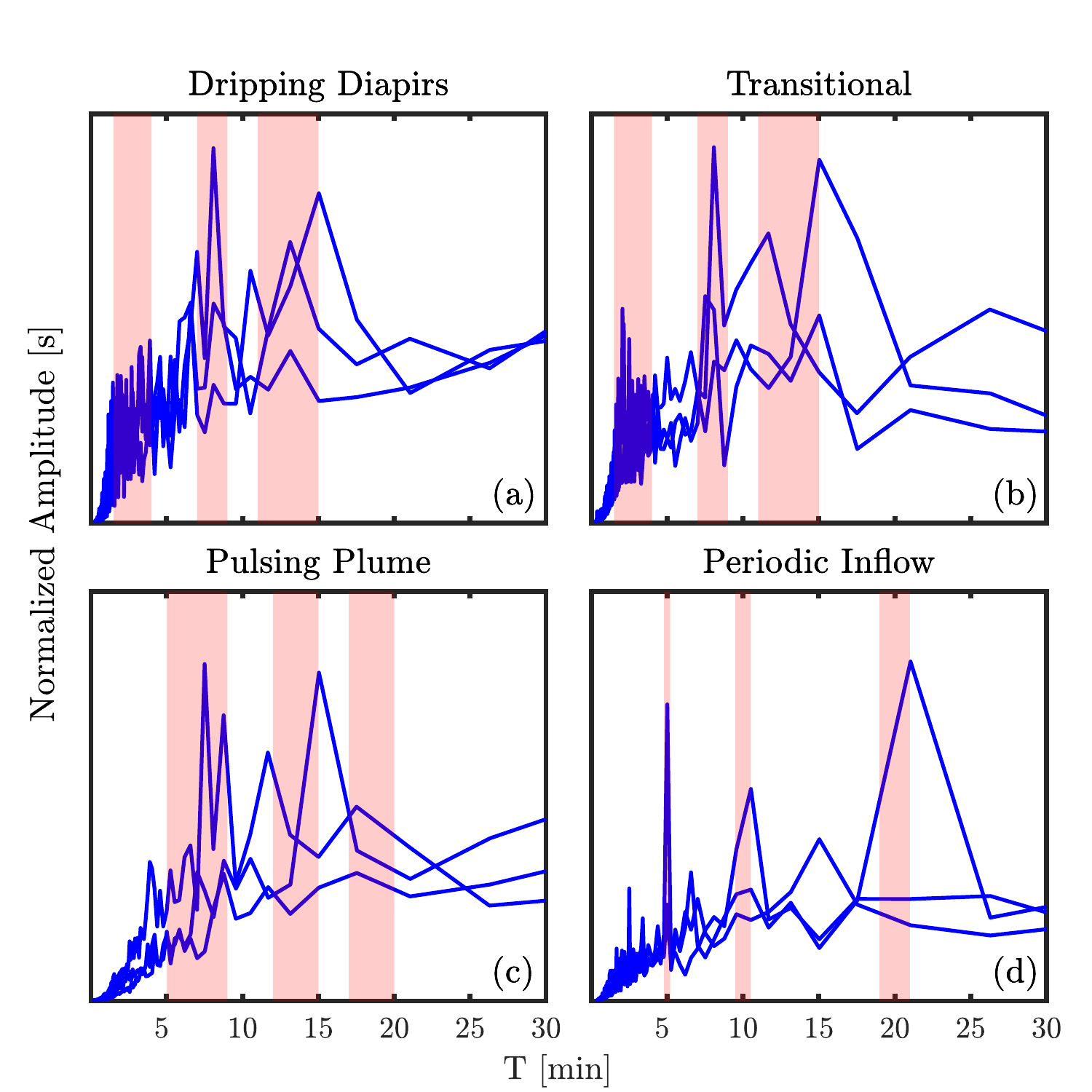}\\
\caption{Surface velocity spectra of additional model runs in the dripping diapirs (a), transitional (b), pulsing plumes (c) regimes, and with periodic inflow conditions (d) in blue. With dominant period ranges highlighted in red. Specific periodicities vary between model runs, even within each regime, but show broadly similar trends corresponding to dripping instabilities at shorter periods and lake reorganization at long periods. Periodic inflow is recoverable, but modified by periods controlled by lake convection.}
\label{fig:Velocity_Fourier_Supplement}
\end{center}
\end{figure}

We include a table of mineral densities used to compute the average crystal density (Table \ref{tbl:MineralDensity}). Functional fit to laboratory measurements of equilibrium crystallinity (Fig.~\ref{fig:Crystallinity}). Plot of viscosity as a function of temperature and crystallinity (Fig.~\ref{fig:Viscosity}). Temperature change over a model run showing heat loss through conduction through the lake walls and cooling of the lake corners with constant lake temperature(Fig.~\ref{fig:Heat_Loss}). Additional parameter variations that modify cooling (Fig.~\ref{fig:Cooling_Variations}), outgassing (Fig~\ref{fig:Outgassing_Variations}), and inflow (Fig.~\ref{fig:Inflow_Variations}) conditions and magma rheology (Fig.~\ref{fig:Rheology_Variations}). Resolution testing showing consistent behavior at higher resolution (Fig.~\ref{fig:Resolution_testing}). Additional selected Fourier analyses in the dripping, transitional, and pulsing regimes and for periodic inflow conditions (Fig.~\ref{fig:Velocity_Fourier_Supplement}). \par
 
\begin{table}[hbp]
\begin{center}
\begin{tabular}{l c c}
Mineral & Volume Fraction & Density\ (kg/m$^3$)\\
\hline\hline
Anorthoclase feldspar & 0.91 & 2580\\
Titanomagnetite & 0.03 & 5180\\
Olivine & 0.02 & 3800\\
Clinopyroxene & 0.02 & 3200\\
Fluorapatite & 0.02 & 3200\\
\hline
Average & & 2720\\
\hline
\end{tabular}
\caption{Mineral densities for the calculation of weighted average crystal density. The crystal volume fractions are an average of measurements from \citet{Moussallam2013}, and density of mineral phases from \citet{Klein2002}.}
\label{tbl:MineralDensity}
\end{center}
\end{table}

\end{document}